\documentclass{article}

\usepackage{PRIMEarxiv}

\usepackage[utf8]{inputenc} 
\usepackage[T1]{fontenc}    
\usepackage{hyperref}       
\usepackage{url}            
\usepackage{booktabs}       
\usepackage{amsfonts}       
\usepackage{nicefrac}       
\usepackage{microtype}      
\usepackage{lipsum}         
\usepackage{mdframed}
\usepackage{graphicx}
\usepackage{natbib}
\usepackage{doi}
\usepackage[table,xcdraw,svgnames]{xcolor}
\usepackage{threeparttable}
\usepackage{algorithm}
\usepackage{algpseudocode}
\usepackage{subcaption}
\usepackage{amsmath}
\usepackage{cleveref} 

\title{Adaptive Plan-Execute Framework for Smart Contract Security Auditing
}

\author{
  Zhiyuan Wei \\
  Beijing Institute of Technology \\
  Beijing, China\\
  \texttt{weizhiyuan@bit.edu.cn} \\
   \And
  Jing Sun \\
  University of Auckland \\
  Auckland, New Zealand\\
  \texttt{jing.sun@auckland.ac.nz} \\
  \And
    Zijian Zhang \\
  Beijing Institute of Technology \\
  Beijing, China\\
  \texttt{zhangzijian@bit.edu.cn} \\
    \And
    Zhe Hou \\
  Griffith University \\
  Queensland, Australia\\
  \texttt{Z.hou@griffith.edu.au} \\
    \And
    Zixiao Zhao \\
  University of Auckland \\
  Auckland, New Zealand\\
  \texttt{zzha584@aucklanduni.ac.nz} \\
}

\begin{document}
\maketitle

\begin{abstract}
  Large Language Models (LLMs) have demonstrated significant potential in smart contract auditing. However, they are still susceptible to hallucinations and limited context-aware reasoning. In this paper, we propose SmartAuditFlow, a dynamic Plan-Execute framework that customizes audit strategies based on the unique characteristics of each smart contract. Unlike static, rule-based workflows, our approach iteratively generates and refines audit plans in response to intermediate outputs and newly detected vulnerabilities. To improve reliability, the framework incorporates structured reasoning, prompt optimization, and external tools such as static analyzers and Retrieval-Augmented Generation (RAG). This multi-layered design reduces false positives and enhances the accuracy of vulnerability detection. Experimental results show that SmartAuditFlow achieves 100\% accuracy on common vulnerability benchmarks and successfully identifies 13 additional CVEs missed by existing methods. These findings underscore the framework’s adaptability, precision, and practical utility as a robust solution for automated smart contract security auditing. 
\end{abstract}


\section{Introduction}
Smart contracts, foundational to blockchain networks and decentralized finance (DeFi), have frequently exhibited critical security vulnerabilities, posing significant financial and operational risks. The inherent complexity of smart contract security is a challenge, particularly as developers may lack comprehensive means to assess security rigorously prior to deployment. Unlike traditional software, where post-deployment patching is standard, the immutable or costly-to-modify nature of deployed smart contracts means that vulnerabilities can lead to substantial and often irreversible losses across temporal, operational, financial, and reputational dimensions. Several high-profile incidents have underscored these vulnerabilities, such as the DAO Exploit (50 million USD), Poly Network Exploit (610 million USD), and Bybit Crypto Exchange Hack (1.5 billion USD). In 2024 alone, financial losses from smart contract vulnerabilities exceeded \textbf{\$2.6 billion} USD across \textbf{192} incidents, resulting from smart contract-level vulnerabilities \footnote{https://getfailsafe.com/failsafe-web3-security-report-2025/}. These incidents underscore the critical and urgent need for robust security auditing mechanisms.

Existing smart contract security auditing primarily relies on manual code reviews by security experts, often supplemented by automated analysis tools \cite{zhang2023demystifying}. While valuable, these approaches face significant limitations in the fast-paced blockchain environment. Manual audits, despite their potential for depth, are inherently time-consuming and costly, often creating bottlenecks in development cycles. Furthermore, the continuous evolution of blockchain technology and the emergence of novel attack vectors challenge the ability of human auditors to consistently identify all vulnerabilities \cite{ sun2024gptscan}.  Expert reviews also rely on individual experience and knowledge, which can introduce subjectivity and gaps in coverage. Conversely, conventional automated tools, while offering speed, frequently suffer from high false positive rates and struggle to detect complex, context-dependent vulnerabilities. Such tools often lack the nuanced understanding of business logic and intricate blockchain-specific security patterns required to identify subtle flaws, such as sophisticated logic errors or vulnerabilities arising from inter-contract interactions, thereby limiting their overall effectiveness.

The advent of Large Language Models (LLMs) presents a promising avenue for advancing smart contract auditing and addressing the aforementioned limitations \cite{liu2024propertygpt}. LLMs possess the unique ability to process and understand both natural language (e.g., contract specifications, comments) and programming languages (i.e., smart contract code) \cite{liu2023your}. This facilitates a deeper reasoning about the alignment between a contract's intended functionality and its actual implementation. Their notable capabilities in pattern recognition and contextual analysis are particularly valuable for detecting subtle or novel security vulnerabilities that traditional automated tools might overlook. Furthermore, LLMs are good at reasoning processes and explaining identified security issues in an accessible manner, thereby empowering developers who may not be security specialists to understand and address the findings.

Recent research has shown encouraging results in applying LLMs to smart contract security analysis \cite{david2023you, wei2023survey}.  Nevertheless, substantial challenges hinder the development of fully reliable and effective LLM-based auditing systems. Key issues include: (1) output inconsistency and non-determinism \cite{bender2021dangers}; (2) the difficulty of validating the correctness and completeness of LLM-generated security assessments; and (3) limitations in ensuring comprehensive exploration of potential vulnerabilities. 
These factors necessitate careful strategies to harness LLM strengths while mitigating their inherent weaknesses. These challenges can be effectively addressed by adopting a \textbf{workflow-based} approach. This methodology decomposes the complex auditing task into manageable, discrete steps, facilitating structured and iterative interactions with LLMs. The efficacy of such structured methods is evident in the improved performance of advanced LLMs in complex reasoning when guided by systematic procedures \cite{zhong2024evaluation,guan2024deliberative}, and is central to agentic workflows like HuggingGPT, Manus, and AutoGPT \cite{shen2023hugginggpt, wang2024survey}, which orchestrate LLMs for sophisticated problem-solving.

In this paper, we introduce SmartAuditFlow, a novel framework that integrates dynamic audit plan generation with workflow-driven execution for robust smart contract auditing. It systematically structures the auditing process by (1) in-depth analysis of the smart contract’s structural and functional characteristics, (2) generation of an adaptive audit plan, (3) iterative execution of this plan via a multi-step workflow, and (4) progressive refinement of security assessments. By emulating the methodical approach and adaptability of expert human auditing teams, SmartAuditFlow aims to significantly enhance the accuracy, efficiency, and comprehensive security analysis. Our primary contributions are as follows:

\begin{itemize}
    \item \textbf{Dynamic and Adaptive Audit Plan Generation:}  We propose a novel framework for generating audit plans that are dynamically tailored to the specific characteristics (e.g., complexity, potential risk areas identified through static analysis) of each smart contract. This plan adaptively refines its audit strategies based on intermediate findings from LLM analysis, leading to a more targeted and exhaustive security assessment.
    \item \textbf{Iterative Prompt Optimization:}  
    We introduce a dynamic prompt engineering strategy where prompts are iteratively refined throughout the audit lifecycle. This refinement is based on real-time analytical feedback and the evolving audit context, enhancing LLM accuracy, improving the reliability of security assessments, and mitigating common issues such as output misinterpretation or factual hallucination.
    \item \textbf{Integration of External Knowledge Sources:}  
    SmartAuditFlow incorporates a hybrid approach by integrating inputs from traditional static analysis tools (for extracting key code attributes and potential hotspots) and Retrieval-Augmented Generation (RAG). The RAG component leverages external knowledge bases to provide LLMs with up-to-date, authoritative security information and vulnerability patterns, thereby enriching contextual understanding and improving the accuracy and completeness of detected vulnerabilities.
    \item \textbf{Rigorous Evaluation and Efficiency Improvements:} We conduct a comprehensive empirical evaluation of SmartAuditFlow across a diverse set of smart contracts using dvanced performance metrics, including Top-$N$ Accuracy, Mean Reciprocal Rank (MRR), and Mean Average Precision (MAP).  Our results demonstrate that SmartAuditFlow outperforms traditional static analysis tools and existing LLM-based auditing frameworks, detecting a broader spectrum of vulnerabilities while significantly improving computational efficiency and scalability.
\end{itemize}

By integrating adaptive planning with workflow-driven execution, SmartAuditFlow sets a new standard for LLM-powered smart contract security auditing, enabling more precise, efficient, and scalable vulnerability detection.

The rest of the paper is organized as follows. 
Section \ref{sec:backgrounds} presents the current challenges in smart contract analysis and reviews existing methods.
Section \ref{sec:methodology} provides an in-depth explanation of SmartAuditFlow's architecture and operational mechanisms.
Section \ref{sec:experiments} presents a comprehensive evaluation of SmartAuditFlow, including the datasets used, evaluation criteria, experimental results.
Section \ref{sec:related} discusses related works, the findings and examines potential threats to validity.
Section \ref{sec:conclusion} concludes the paper by summarizing its contributions and offering insights into the future of smart contract analysis tools.

\algrenewcommand{\algorithmiccomment}[1]{\hfill\textcolor{gray}{\(\triangleright\) #1}}

\section{Backgrounds}
\label{sec:backgrounds}
\subsection{Smart Contract Auditing}
\label{sec:sca_importance}
The rapid growth of smart contracts across blockchain networks has brought transformative capabilities, particularly to the Decentralized Finance (DeFi) sector. This rapid adoption, however, has simultaneously highlighted the critical importance of robust security measures, as vulnerabilities within smart contracts have frequently led to substantial financial losses \cite{tolmach2021survey, wei2023survey}. 
The DeFi market, for instance, reached a peak total value locked (TVL) of approximately \$179 billion USD on November 9, 2021, with the Ethereum network alone contributing \$108.9 billion USD (around 60\% of the total DeFi TVL at the time). 
As these blockchain-based applications attract significant capital and user engagement, they have also become prime targets for malicious actors. Until the end of 2024, the total amount of money lost by blockchain hackers exceeds 35.32 billion USD, stemming from more than 1800 hacks. A particularly severe recent incident occurred in February 2025, involving the theft of approximately \textbf{\$1.5 billion} USD in Ethereum from the Bybit digital asset exchange, reportedly one of the largest single cryptocurrency heists to date \footnote{https://www.csis.org/analysis/bybit-heist-and-future-us-crypto-regulation}.

A smart contract audit is a meticulous and systematic security assessment designed to identify and mitigate vulnerabilities, logical errors, and inefficiencies within the contract's codebase. Different from foundational blockchain security concerning consensus mechanisms or virtual machines, smart contract security operates at the application layer, which may be more irregular and complex. 
During an audit, security experts scrutinize all pertinent materials, including source code, whitepapers, and architectural documentation. 
This process typically involves manual inspection to uncover nuanced flaws that automated tools might miss, complemented by the use of specialized static and dynamic analysis tools to detect common vulnerability patterns and deviations from best practices \cite{ma2024combining, xia2024auditgpt}. The cost of such audits can range from \textbf{\$5,000} to over \textbf{\$50,000} USD, contingent on contract complexity, and demand deep expertise in software engineering, blockchain-specific programming languages (like Solidity), and DeFi protocols. 
Audit durations can vary from a few days for simple contracts to several weeks or months for complex decentralized applications (dApps).

\subsection{Automated Vulnerability Detection}
Traditional automated techniques for smart contract vulnerability detection, including static analysis, symbolic execution, and fuzz testing, have been instrumental in early efforts to secure smart contracts \cite{chaliasos2024smart, durieux2020empirical}. However, these methods face significant challenges related to scalability, computational efficiency, and their ability to comprehensively analyze increasingly complex smart contract systems. 
A critical study by Zhang et al. \cite{zhang2023demystifying} revealed that 37 widely-used automated detection tools failed to identify over 80\% of 516 exploitable vulnerabilities found in 167 real-world smart contracts, underscoring substantial blind spots in conventional approaches. The study also highlighted that the intricacies of detecting complex bugs often necessitate the collaborative efforts of multiple human auditors, further emphasizing the limitations of standalone automated tools and the high cost of thorough manual reviews.

Recent advancements in LLMs offer a promising frontier for enhancing smart contract security analysis \cite{david2023you, wei2023survey}. LLMs demonstrate strong capabilities in understanding and reasoning about both natural language and programming code. 
This allows them to potentially bridge the gap between intended contract behavior and actual implementation logic, identifying vulnerabilities, verifying compliance with specifications, and assessing logical correctness with a degree of nuance previously unattainable by purely algorithmic tools. 
LLMs can excel at reasoning about complex security scenarios, explaining identified vulnerabilities in an understandable manner, evaluating business logic intricacies, and even hypothesizing novel attack vectors \cite{chen2023chatgpt, sun2024gptscan}. 
Studies exploring LLMs in the smart contract domain have reported promising results, in some instances demonstrating performance comparable or superior to traditional automated tools. However, the probabilistic nature of LLMs means that their outputs can be inconsistent, and developing robust mechanisms for validating their findings remains an active area of research and a critical consideration for practical deployment \cite{azamfirei2023large, chen2023diversevul}.

\subsection{Agentic Systems and Workflows}
Effectively harnessing LLMs for complex tasks like smart contract auditing requires sophisticated interaction strategies beyond simple querying.
Prompt engineering, the art of crafting effective inputs to guide LLM outputs, is a cornerstone of LLM application. It is often more computationally efficient and cost-effective for adapting pre-trained LLMs to specific tasks than full model retraining or extensive fine-tuning, especially for rapid prototyping and iteration. 
However, LLMs are highly sensitive to prompt phrasing; minor variations in input can lead to significantly different, sometimes degraded, performance. For instance, one study demonstrated accuracy drops of up to \textbf{65\%} across several advanced LLMs due to subtle changes in question framing \cite{mirzadeh2024gsm}. To address this sensitivity and enhance LLM reasoning, researchers have developed advanced prompting methodologies. These include techniques like chain-of-thought (CoT) prompting, which encourages step-by-step reasoning \cite{li2023camel}; few-shot learning, which provides examples within the prompt \cite{wang2024chain}; and more complex structured reasoning frameworks like Tree of Thought (ToT) \cite{yao2024tree} and Graph of Thought (GoT) \cite{besta2024graph}. These methods aim to elicit more reliable and accurate responses by structuring the LLM's generation process.

Beyond one attempt advanced prompting, more sophisticated operational paradigms are emerging to tackle multi-step, complex problems. Two prominent approaches are autonomous agents and agentic workflows.
\textbf{Autonomous agents} are designed to perceive their environment, make decisions, and take actions dynamically to achieve goals, often involving flexible, self-directed reasoning and tool use \cite{zhugegptswarm}. They typically require careful design of their action space and decision-making heuristics tailored to their operational context.
In contrast, \textbf{agentic workflows} structure problem-solving by decomposing a complex task into a sequence of predefined or dynamically chosen sub-tasks, often involving multiple LLM invocations, potentially with specialized roles or prompts for each step \cite{hong2024data, zhang2024aflow}. This paradigm emphasizes a more explicit orchestration of LLM capabilities, leveraging human domain expertise in designing the workflow structure and allowing for iterative refinement of both the process and the intermediate results. Agentic workflows can integrate external tools, knowledge bases, and feedback loops, making them well-suited for tasks requiring methodical execution and high reliability, such as security auditing.

\section{Methodology}
\label{sec:methodology}
In this section, we present the design and operational principles of SmartAuditFlow, our LLM-powered framework for automated smart contract auditing. We begin with an overview of the framework's architecture and its core Plan-Execute reasoning paradigm (Section \ref{subsec:overview}). Subsequently, we will discuss in detail the primary operational phases: the Planning Phase, encompassing initial analysis and audit strategy generation (detailed in Section \ref{subsec:planning_phase_details}), and the Execution Phase, covering multi-faceted vulnerability assessment and synthesis (detailed in Section \ref{subsec:execute-phase}). Finally, we design a LLM-Powered Audit Evaluator to measure the quality of the audit report (Section \ref{subsec:llm_evaluator}).

\begin{figure}[t]
	\centering
	\includegraphics[width=1\linewidth]{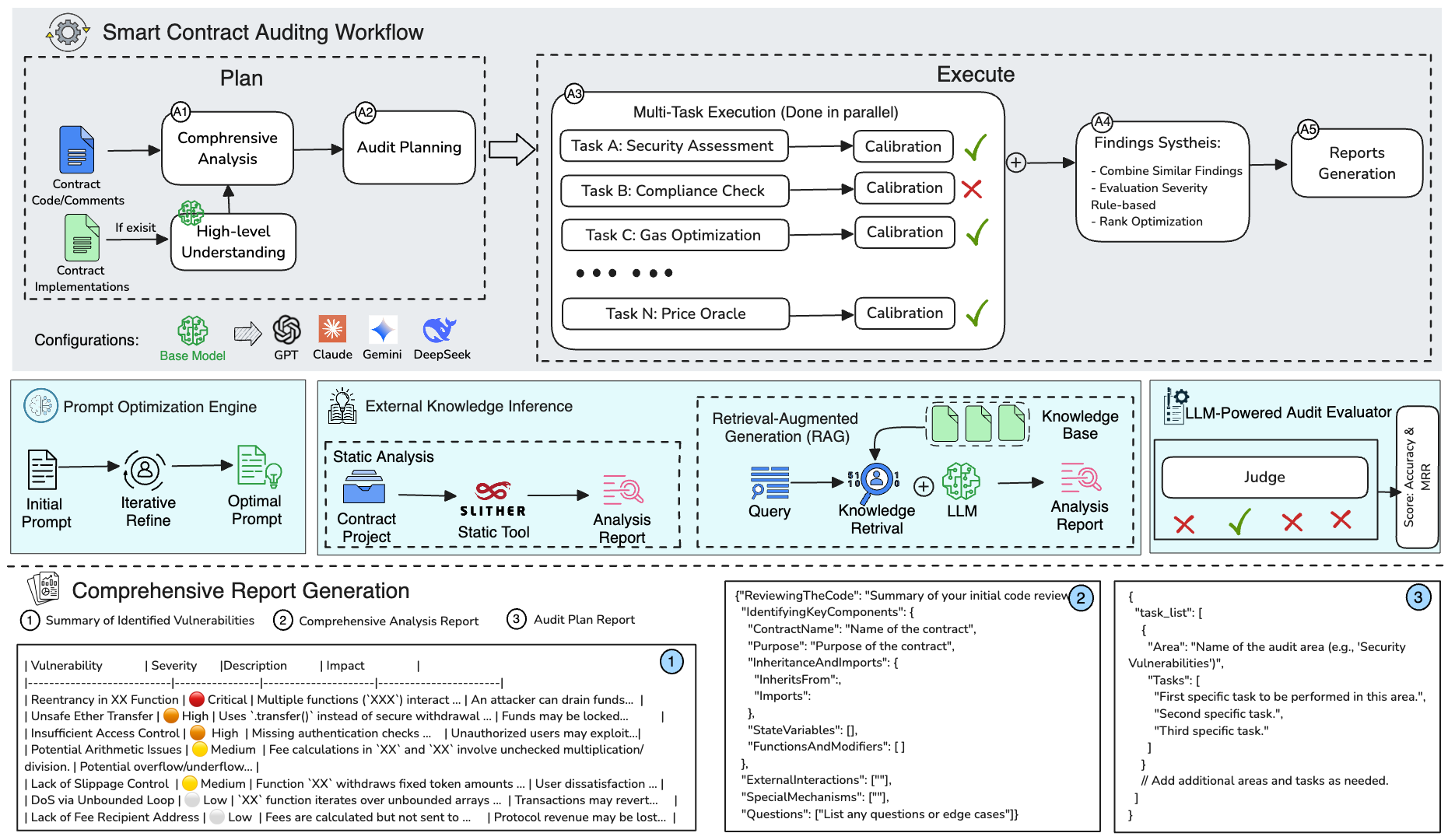}
	\caption{Overview of the SmartAuditFlow system.}
	\label{fig:smartauditflow_stages}
\end{figure}

\subsection{Overview}
\label{subsec:overview}
To address the inherent complexities of smart contract auditing, the uncertainties in LLM-based code analysis, and the significant workload of human auditors, SmartAuditFlow employs a structured \textbf{Plan-Execute Reasoning} paradigm. This paradigm decomposes the audit into five sequential stages, each driven by an LLM and tailored to a distinct aspect of the audit, ensuring a systematic, coherent, and actionable review.
The overarching aim is to meticulously analyze smart contracts and generate a comprehensive, coherent, and actionable audit report.

Figure~\ref{fig:smartauditflow_stages} depicts the high-level architecture of SmartAuditFlow, which is composed of five main components. The \textit{Smart Contract Auditing Workflow} serves as the central processing pipeline and its operational logic is formally defined in Algorithm~\ref{alg:smartauditflow}. Supporting this workflow, the \textit{Prompt Optimization Engine} is dedicated to the generation, refinement, and validation of optimal prompts, which are crucial for guiding the LLM effectively at each stage. Furthermore, \textit{External Knowledge Inference} augments the LLM's inherent knowledge and improves analytical accuracy by incorporating external data sources such as up-to-date vulnerability databases and security best practices. To ensure the reliability of the audit, the \textit{LLM-Powered Audit Evaluator} establishes a framework for fair, consistent, and reproducible evaluations of the audit outputs generated by the LLM judge. Finally, the \textit{Comprehensive Report Generation} component is responsible for consolidating all validated findings, analyses, and supporting evidence from the Core Auditing Workflow into a structured, actionable, and clear audit report suitable for various stakeholders.

The heart of SmartAuditFlow's operation is in its Core \textit{Auditing Workflow}, which is precisely delineated in Algorithm~\ref{alg:smartauditflow}. This algorithm outlines a dynamic, multi-stage process for auditing smart contracts.  It employs an LLM ($\mathcal{M}$), guided by a set of specialized prompts ($\mathcal{P}$), takes smart contract code ($c$) and contextual documents ($D_{ctx}$) as primary inputs, and produces a comprehensive audit report ($R$).

Initially, the LLM leverages prompt $\mathcal{P}_{A1}$ with $c$ and $D_{ctx}$ to generate an initial contextual understanding ($s_1$). Subsequently, this understanding $s_1$ informs the LLM, via prompt $\mathcal{P}_{A2}$ and code $c$, in creating a prioritized audit plan ($s_2$), which is then decomposed into specific sub-tasks ($\mathbf{t}$). Each sub-task $t^j$ then undergoes a multi-faceted execution and validation process; here, the LLM, using task-specific execution ($\mathcal{P}_{A3e}^j$) and validation ($\mathcal{P}_{A3v}^j$) prompts, first performs a preliminary assessment ($e^j$) and then validates this to yield a finding $v^j$ with a confidence score, aggregating only those findings that meet a predefined threshold into the set of validated findings ($s_3$). These validated findings $s_3$ are then subjected to a cross-cutting synthesis, producing a synthesized security posture overview ($s_4$). The entire workflow culminates in the generation of the final audit report ($R$), which contains the initial understanding $s_1$, the audit plan $s_2$, and the synthesized findings $s_4$.

\begin{algorithm}[t]
	\caption{Dynamic Smart Contract Auditing Workflow}
	\label{alg:smartauditflow}
	\begin{algorithmic}[1]
	\Require{Smart contract code $c$, Contextual documents $D_{ctx}$, LLM $\mathcal{M}$, 
  \Statex Set of specialized prompts $\mathcal{P} = \{\mathcal{P}_{A1}, \mathcal{P}_{A2}, \{\mathcal{P}_{A3e}^i\}, \{\mathcal{P}_{A3v}^i\}, \mathcal{P}_{A4}\}$}
	\Ensure{Comprehensive audit report $R$}
  \Statex \textit{\textcolor{gray!60}{// A1: Context-Aware Initial Analysis}}
  \State $s_1 \gets \mathcal{M}(\mathcal{P}_{A1}(c, D_{ctx}))$ \Comment{Generate initial analysis and understanding}

  \Statex \textit{ \textcolor{gray!60}{// A2: Adaptive Audit Planning}}
  \State $s_2 \gets \mathcal{M}(\mathcal{P}_{A2}(s_1, c))$ \Comment{Create a prioritized audit plan (list of sub-tasks $\mathbf{t}$)}
  \State $\mathbf{t} = \{t^1, t^2, \dots, t^n\} \gets \text{extract\_subtasks}(s_2)$

  \Statex \textit{\textcolor{gray!60}{// A3: Multi-faceted Vulnerability Execution and Validation}}
  \State $S_v = \emptyset$ \Comment{Initialize set of validated findings for $s_3$}
  \For{each sub-task $t^j \in \mathbf{t}$}
    \State $e^j \gets \mathcal{M} (\mathcal{P}_{A3e}^j(t^j, c, s_1))$ \Comment{Preliminary execution for sub-task $t^j$}
    \State $(v^j, \text{confidence}^j) \gets \mathcal{M}(\mathcal{P}_{A3v}^j(e^j, c))$ \Comment{Validate $e^j$; output validated finding $v^j$ and confidence}
    \If{$\text{confidence}^j \geq \text{THRESHOLD\_CONFIDENCE}$}
      \State $S_v \gets S_v \cup \{ v^j \}$
    \EndIf
  \EndFor
  \State $s_3 \gets S_v$ \Comment{Aggregated validated findings from execution stage}

  \Statex \textit{\textcolor{gray!60}{// A4: Cross-Cutting Findings Synthesis}}
  \State $s_4 \gets \mathcal{M}(\mathcal{P}_{A4}(s_3))$ \Comment{Correlate, prioritize, assign severity to findings in $s_3$}

  \Statex \textit{\textcolor{gray!60}{// A5: Comprehensive Report Generation}}
  \State $s_5 \gets \{s_1, s_2, s_4\}$ 
  \State \Return{$R=s_5$}
	\end{algorithmic}
\end{algorithm}

\subsection{Plan Phase}
\label{subsec:planning_phase_details}
The Plan phase is critical for preparing focused and effective tasks for the subsequent Execute Phase. Central to this is guiding the LLM using precisely engineered prompts. Prompts fundamentally shape the LLM's interpretation of smart contract structures, identification of high-risk functions, and overall focus during analysis. As highlighted by prior research, poorly constructed or misaligned prompts can significantly degrade LLM output quality, potentially leading to incomplete audits or misinterpretation of vulnerabilities \cite{zhong2024evaluation, guan2024deliberative}.  Therefore, the systematic generation and selection of optimal prompts for each reasoning step is crucial. 

\subsubsection{Initial Analysis and Adaptive Planning Stages (A1-A2)}
\label{subsubsec:initial_analysis_planning}
The Plan phase comprises two foundational stages: Context-Aware Initial Analysis (A1) and Adaptive Audit Planning (A2). The audit starts with a thorough understanding $s_1$ of the smart contract project. The objectives at this stage are to interpret the contract's business logic, identify its main roles (e.g., token standard, governance, upgradeability, DeFi interactions), delineate key functionalities, and recognize potentially exposed assets (e.g., funds, sensitive data, privileged permissions).
For instance, understanding the design standards for specific token types (e.g., ERC-20, ERC-721) and their common implementation patterns is essential. The LLM systematically analyzes the contract’s structure and logic, methodically examining functions and data structures to highlight preliminary areas of interest or potential high-level concerns.

In Stage A2, the LLM leverages insights from the initial analysis ($s_1$) to formulate a tailored audit plan ($s_2$). 
This plan outlines a structured set of sub-tasks, each targeting specific security aspects of the contract, such as reentrancy, access control, or token issue.
The plan is designed to focus analytical efforts on the most critical areas first. The quality of the audit plan depends heavily on the LLM’s interpretation of $s_1$. Additionally, the LLM may be prompted to identify complex functions or contract sections, recommending further decomposition for granular analysis in subsequent stages. This adaptive process ensures the audit strategy is customized for each contract's unique risk profile.

To achieve A1 and A2 stages, we employ an iterative prompt optimization strategy inspired by recent advancements in automatic prompt engineering \cite{weng2022large, zhou2022large, cheng2023black}. This strategy aims to discover an optimal instruction $\rho^*$ that, when prepended to a given smart contract code $c$, maximizes the quality of the LLM's output $A$ (e.g., a comprehensive initial analysis or a robust audit plan). 
Moreover, Stage A1 is enhanced by integrating insights from static analysis tools to ground the LLM's assessments and improve the accuracy of its initial findings. These insights inform the subsequent prompt generation and planning in later stages.

\begin{algorithm}[htp]
  \caption{Iterative Prompt Optimization Algorithm}
  \label{alg:genetic_algorithm}
  \begin{algorithmic}[1]
  \Require Training dataset $\mathcal{D}_{train}$, validation set $\mathcal{D}_{val}$;
  \Statex Max generations $T_{\text{max}}$, population size $P_S$, elite count $k_e$, offspring count $k_o = P_S - k_e$; 
  \Statex Replay buffer capacity $m$, mutation parameters $(\tau_{\max}, \beta)$
  \Statex Convergence criteria: min fitness improvement $\delta_{\text{fitness}}$, stable generations $N_{\text{stable}}$, min diversity $D_{\text{min}}$
  \Ensure Optimal instruction $\rho^*$

  \Statex \textbf{Phase 1: Initialization:}
  \State Generate initial population $\mathcal{U}_0 = \{\rho_1, \rho_2, \dots, \rho_{P_S}\}$ using mutation of seed prompts
  \State Initialize replay buffer $\mathcal{B}_{\text{replay}} \gets \emptyset$
  \State Initialize $\bar{f}_{0}(\rho) \gets \text{EvaluateInitialFitness}(\rho, \mathcal{D}_{\text{train}})$ for each $\rho \in \mathcal{U}_0$ 
  \State $t \gets 1$; $generations\_without\_improvement \gets 0$
  
  \Statex \textbf{Phase 2: Evolutionary Loop}
  \While{$t \leq T_{\text{max}}$ \textbf{and} $generations\_without\_improvement < N_{\text{stable}}$}
    
      \Statex \textit{\hspace{\algorithmicindent} \textcolor{gray!60}{// A. Mini-Batch Sampling}}
      \State Sample mini-batch $\mathcal{B}_t \subset \mathcal{D}_{train}$ of size $n_t = \lceil 0.1|\mathcal{D}_{train}| \cdot (1 + t/T) \rceil$
      
      \Statex \textit{\hspace{\algorithmicindent} \textcolor{gray!60}{// B. Stochastic Fitness Evaluation for Population $\mathcal{U}_{t-1}$}}
      \ForAll{$\rho \in \mathcal{U}_{t-1}$}
          \If{$|\mathcal{B}_{\text{replay}}| > 0$}
          \State $f_{\text{replay}}(\rho) \gets \epsilon \cdot \frac{1}{|\mathcal{B}_{\text{replay}}|} \sum_{(\rho', f') \in \mathcal{B}_{\text{replay}}} \text{sim}(\rho, \rho') \cdot f'$ 
          \Else
            \State $f_{\text{replay}}(\rho) \gets 0$
          \EndIf
           \State $f_{\text{batch}}(\rho) \gets \frac{1}{|\mathcal{B}_t|} \sum_{(c,A) \in \mathcal{B}_t} f(\rho, c, A)$
          \State $f_t(\rho) \gets f_{\text{batch}}(\rho) + f_{\text{replay}}(\rho)$
      \EndFor
      
      \Statex \textit{\hspace{\algorithmicindent} \textcolor{gray!60}{// C. Update Smoothed Fitness and Select Elites from $\mathcal{U}_{t-1}$}}
      \ForAll{$\rho \in \mathcal{U}_{t-1}$}
        \State $\bar{f}_t(\rho) \gets \alpha \cdot \bar{f}_{t-1}(\rho) + (1 - \alpha) \cdot f_t(\rho)$
      \EndFor
      \State Select $k_e$ elites $\mathcal{E}_t$ based on $\bar{f}_t(\cdot)$

      \Statex \textit{\hspace{\algorithmicindent} \textcolor{gray!60}{// D. Offspring Generation}}
      \State Compute mutation temperature: $\tau_t \gets \tau_{\text{max}} \cdot e^{-\beta t}$
      \State Select parents $\mathcal{P}_{\text{parents}}$ from $\mathcal{U}_{t-1}$ (e.g., using tournament selection, or use $\mathcal{E}_t$)
      \State $\mathcal{O}_t \gets \emptyset$
      \For{$i=1$ \textbf{to} $k_o$} \Comment{Generate $k_o = P_S - k_e$ offspring}
          \State $\rho_{\text{parent}} \gets \text{SelectParent}(\mathcal{P}_{\text{parents}})$
          \State $\rho_{\text{offspring}} \gets \text{Mutate}(\rho_{\text{parent}}, \tau_t, \mathcal{B}_t)$ \Comment{Guided mutation using $\mathcal{B}_t$}
          \State $\mathcal{O}_t \gets \mathcal{O}_t \cup \{\rho_{\text{offspring}}\}$
          \State $\bar{f}_{t-1}(\rho_{\text{offspring}}) \gets \bar{f}_{t-1}(\rho_{\text{parent}})$ \Comment{Initialize smoothed fitness for new offspring, or use parent's}
      \EndFor

      \Statex \textit{\hspace{\algorithmicindent} \textcolor{gray!60}{// E. Form New Population for Next Generation}}
      \State $\mathcal{U}_t \gets \mathcal{E}_t \cup \mathcal{O}_t$ \Comment{New population of size $P_S = k_e + k_o$}
      \Statex \textit{\hspace{\algorithmicindent} \textcolor{gray!60}{// F. Update Replay Buffer}}
      \State Update $\mathcal{B}_{\text{replay}}$ with top-$m$ performers from $\mathcal{U}_t$

      \Statex \textit{\hspace{\algorithmicindent} \textcolor{gray!60}{// G. Convergence Check}}
      \State $\bar{f}_{\text{best}, t} \gets \max_{\rho \in \mathcal{U}_t} \bar{f}_t(\rho)$
      \If{$t > 1$ \textbf{and} $|\bar{f}_{\text{best}, t} - \bar{f}_{\text{best}, t-1}| < \delta_{\text{fitness}}$}
          \State $generations\_without\_improvement \gets generations\_without\_improvement + 1$
      \Else
          \State $generations\_without\_improvement \gets 0$
      \EndIf
      \State $D(\mathcal{U}_t) \gets 1 - \text{MeanPairwiseSimilarity}(\mathcal{U}_t)$ 
      \State $t \gets t + 1$
  \EndWhile
  
  \Statex \textbf{Phase 3: Final Selection}
  \State Using Eq. \ref{eq:optimization} to select $\rho^*$ on a held-out validation set \(\mathcal{D}_{\text{val}}\)
  \State \Return $\rho^*$
  \end{algorithmic}
\end{algorithm}

\subsubsection{Iterative Prompt Optimization Framework}
\label{subsubsec:iterative_prompt_opt}
In the Plan Phase, each prompt $\mathcal{P}$ is composed of an instruction component $\rho$ and the smart contract code $c$. Given that $c$ is fixed for a specific auditing task, our optimization focuses on refining the instruction $\rho$. Formally, consider a task specified by a dataset $\mathcal{D}_{task} = \{(c, A)\}$ of input/output pairs, where $c$ is a contract and $A$ is the expected result (e.g., an expert-crafted initial analysis or audit plan).
Our goal is to find an optimal instruction $\rho^*$ that maximizes a scoring function $f(\rho, c, A)$, which measures the alignment of the LLM's output with $A$. We formulate this objective as the following optimization problem:

\begin{equation} 
  \rho^{*} = \arg \max_{\rho} \left( \frac{1}{|\mathcal{D}_{train}|} \sum_{(c,A) \in \mathcal{D}_{train}} f(\rho, c, A) - \lambda \cdot \text{Complexity}(\rho) \right)
  \label{eq:optimization} 
\end{equation}
where $f(\rho, c, A)$ is the multi-criteria scoring function, $\text{Complexity}(\rho)$ measures the complexity of the instructions, $\lambda$ is a regularization hyperparameter, and the maximization is performed over a held-out validation set $\mathcal{D}_{\text{train}}$.

We employ an evolutionary algorithm to search for $\rho^*$, outlined in Algorithm~\ref{alg:genetic_algorithm}. This iterative process is designed to converge towards an instruction that elicits accurate, thorough, and contextually relevant outputs from the LLM $\mathcal{M}$.

\begin{itemize} 
  \item \textbf{Initial Population Generation $(\mathcal{U}_0)$}: We create a diverse initial population of $k$ candidate instructions $\mathcal{U}_0 = \{\rho_1, \rho_2, \dots, \rho_k\} $ using stochastic sampling and mutation:  
  \begin{enumerate}
    \item \textbf{Stochastic Sampling}: A meta-prompt (see Appendix A.1) guides the LLM to generate varied candidate instructions based on task descriptions and examples of high-performing cases.
    \item \textbf{Mutation of Seed Prompts}: We apply paraphrasing, keyword substitution, and structural edits to existing instructions, controlled by a temperature parameter $\tau$ to balance creativity and determinism.
    \item \textbf{Replay Buffer \(\mathcal{B}_{\text{replay}}\)}: Initialize \(\mathcal{B}_{\text{replay}} = \emptyset\), this buffer will store top instructions from previous generations along with their past fitness scores.
  \end{enumerate}
  \item \textbf{Evolution Loop with Mini-Batch Optimization \cite{li2014efficient, stapor2022mini}}:  
  Let \(\mathcal{U}_{t}\) denote the instruction set at generation $t$. Each iteration of the loop ($t=1 \dots T$) proceeds as follows:
  \begin{enumerate}
    \item \textbf{Mini-Batch Sampling}: For generation \(t\), a mini-batch \(\mathcal{B}_t \subset \mathcal{D}_{{train}}\) is sampled, with its size \(n_t\) (e.g., \(n_t = \lceil 0.1|\mathcal{D}_{{train}}| \cdot (1 + t/T) \rceil\)).  
    \item \textbf{Stochastic Fitness Evaluation}: Evaluate each candidate \(\rho \in \mathcal{U}_{t-1}\) using:  
    \begin{equation}
      f_t(\rho) = \frac{1}{|\mathcal{B}_t|} \sum_{(c,A) \in \mathcal{B}_t} {f(\rho, c, A)} + \epsilon \cdot {\frac{1}{|\mathcal{B}_{\text{replay}}|}\sum_{(\rho', f') \in \mathcal{B}_{\text{replay}}} \text{sim}(\rho, \rho') \cdot f'}
      \label{eq:fintness} 
    \end{equation}
    where $f(\rho, c, A)$ is the primary scoring function (Eq. \ref{eq:scoring_metric}), \(\text{sim}(\rho, \rho')\) measures similarity between instructions (e.g., using sentence embeddings), \(\epsilon\) is a hyperparameter weighting the replay-based regularization, and $f'$ representing the previously recorded fitness for $\rho'$ stored in $\mathcal{B}_{\text{replay}}$.
    \item \textbf{Elite Selection with Momentum}: The top-$k_e$ elite instructions $\mathcal{E}_t$ are selected based on a moving average of their fitness to preserve high-performing traits and reduce noise from mini-batch sampling:
    \[ \bar{f}_t(\rho) = \alpha \cdot \bar{f}_{t-1}(\rho) + (1 - \alpha) \cdot f_t(\rho), \]
    where $\alpha \in [0, 1]$ is the momentum coefficient.
    \item \textbf{Offspring Generation with Guided Mutation}: A new set of offspring instructions $\mathcal{O}_t$ is generated by applying mutation operators to selected parents from $\mathcal{E}_t$. The mutation diversity is controlled via exponential decay $\tau_t = \tau_{\text{initial}} \cdot e^{-\beta t}$. Mutations that yield improved scores on the current mini-batch $\mathcal{B}_t$ can be prioritized or reinforced.
    \item \textbf{Population Update}: The next generation's population $\mathcal{U}_t$ is formed by combining elites and offspring, e.g., $\mathcal{U}_t = \mathcal{E}_t \cup \mathcal{O}_t$. The replay buffer $\mathcal{B}_{\text{replay}}$ is updated with the top-performing instructions from $\mathcal{U}_t$.
    \item \textbf{Convergence Check}: The loop terminates if the improvement in the moving average fitness of the top elites plateaus (e.g., $\Delta \bar{f}_t < \delta$ for several generations) or if population diversity \(D(\mathcal{U}_t) = \frac{1}{k^2} \sum_{\rho_i, \rho_j \in \mathcal{U}_t} \text{sim}(\rho_i, \rho_j)\) falls below a threshold.  

  \end{enumerate}
  \item \textbf{Final Instruction Selection}: Upon termination of the evolutionary loop, the instruction $\rho^*$ from the final population that maximizes the objective function defined in Eq. \ref{eq:optimization} (evaluated on the held-out validation set $\mathcal{D}_{\text{val}}$) is selected as the optimal instruction for the given task.
\end{itemize}

\paragraph{Multi-Criteria Scoring Function $f(\rho, c, A)$} 
A simplistic scoring function (e.g., binary success/failure) may fail to capture nuanced alignment between model outputs and desired results. 
Therefore, we combine multiple criteria (e.g., correctness, efficiency, and user satisfaction) into \(f(\rho, c, A)\):

\begin{equation} 
  f(\rho, c, A) = w_{\text{exec}} \cdot f_{\text{exec}}(\mathcal{M}(\rho, c), A) + w_{\text{log}} \cdot f_{\text{log}}(\mathcal{M}(\rho, c), A) \label{eq:scoring_metric} 
\end{equation}
where $\mathcal{M}(\rho, c)$ is the output generated by the LLM when prompted with instruction $\rho$ and contract $c$. The weights $w_{\text{exec}}$ and $w_{\text{log}}$ (e.g., $w_{\text{exec}} + w_{\text{log}} = 1$) balance the contribution of each component.

\paragraph{Output Alignment Score $(f_{\text{exec}})$:} This component measures how well the LLM's generated output $\mathcal{M}(\rho, c)$ aligns with the expected output $A$ from the dataset $\mathcal{D}_{task}$. For smart contract auditing tasks, $A$ might represent an expert-identified set of vulnerabilities, a correctly summarized contract behavior, or a well-structured audit plan component. Alignment can be measured using metrics like BERTScore, ROUGE, or BLEU \cite{vaswani2017attention, es2024ragas} for textual similarity if $A$ is text, or more domain-specific metrics. 
We define a domain-specific accuracy metric, drawing upon best practices and guidelines from leading smart contract audit firms (e.g., ConsenSys Diligence, Trail of Bits, or OpenZeppelin).

\paragraph{Output Likelihood Score ($f_{\text{log}}$):}  This component uses the negative log-likelihood assigned by the LLM $\mathcal{M}$ to its own generated output. A higher log probability suggests that the output is more confident according to the LLM's internal model. It is typically calculated as the average negative log-likelihood per token of the output sequence $O = w_1, \dots, w_N$ generated by $\mathcal{M}(\rho,c)$:
\begin{equation}  f_{\text{log}}(\mathcal{M}(\rho,c), A) = - \frac{1}{N} \sum_{i=1}^N \log P_{\mathcal{M}}(w_i | \rho, c, w_1, \dots, w_{i-1}) 
\label{eq:log_likelihood}
\end{equation}

Additional details, including further equations and the instance for our initial prompt design, are provided in Appendix~\ref{sec:prompt_initial}, where we elaborate on the scoring function and optimization process.

\subsubsection{Static Analysis Tool Involvement}
While LLMs demonstrate remarkable capabilities in processing natural  language and understanding high-level code logic, they may not consistently identify certain edge-case scenarios or subtle security flaws rooted in complex code interdependencies or low-level implementation details. To address this and enrich the initial stages of our audit workflow, we proposed to incorporate insights from static analysis tools.

Widely recognized tools such as Slither, Mythril, Solint, and Oyente automatically detect common vulnerability patterns. However, their utility in our framework extends beyond standalone detection. Specifically, within Stage A1, the outputs from these tools are leveraged to significantly enhance the LLM's judgements. Among these tools, we selected slither-0.11.0 as our static analysis tool. This choice is motivated by Slither's active maintenance, frequent updates driven by a large developer community, and its comprehensive suite of nearly 100 distinct vulnerability detectors.
In contrast, tools like Mythril primarily focus on low-level EVM bytecode analysis, which, while valuable, offers a different granularity of insight, and other tools such as Solint and Oyente have seen limited updates in recent years. Slither's proficiency in extracting detailed code syntax and semantic information (e.g., inheritance graphs, function call graphs, state machine representations) has been well-documented in prior research \cite{huang2023semantic, ayub2023storage}.

Furthermore, the integration of static analysis serves as a crucial cross-referencing mechanism. By comparing LLM-generated insights about potential vulnerabilities against the findings from static analysis, SmartAuditFlow can mitigate the risk of LLM "hallucinations". This hybrid approach effectively balances automated pattern detection with contextual reasoning.

\begin{figure*}[t]
	\centering 
	\includegraphics[width=0.7\linewidth]{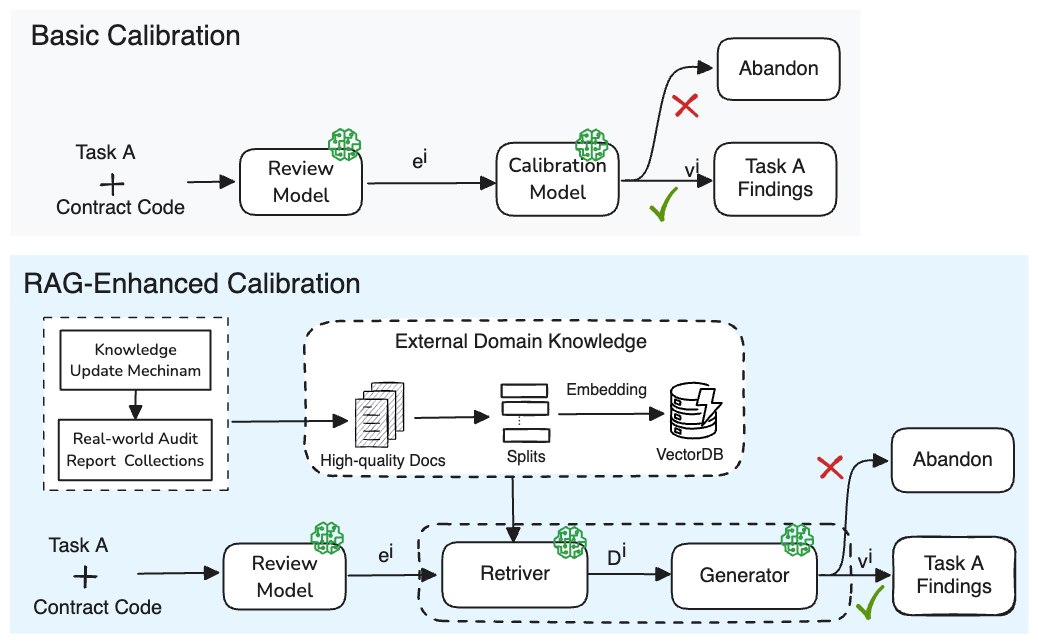}
	\caption{(a) Basic Calibration directly uses LLM to validate the initial review finding; (b) RAG-Enhanced Calibration integrates external knowledge for robust validation and comprises two primary components: Retriever and Generator.}
	\label{fig:rag}
\end{figure*}

\subsection{Execute Phase}
\label{subsec:execute-phase}
The Execute Phase operationalizes the adaptive audit plan by performing detailed, sub-task-specific security checks on the smart contract. 
Unlike the Plan Phase, where prompts are often more extensive to define strategic tasks or perform broad initial analyses, the prompts utilized for individual sub-tasks in the Execute Phase are typically simpler and more direct. Each sub-task is narrowly focused on a single security aspect (e.g., checking for reentrancy in a specific function, evaluating access controls for a particular state variable). This focused approach guides the LLM to conduct a deep exploration of a well-defined issue. 

\subsubsection{Execution and Reporting (A3-A5)}
The Execute Phase begins with the Multi-Task Execution (A3), where the framework executes the prioritized sub-tasks defined in the audit plan ($s_2$). For each sub-task (e.g., "check function X for reentrancy", "analyze access controls for admin functions"), the LLM performs a detailed security assessment on the relevant code segments. This generates preliminary findings. Crucially, these preliminary findings undergo a \textbf{calibration process}. This calibration involves the LLM re-evaluating its own initial finding, potentially guided by additional information such as rule-based checks or specific validations. Only findings $s_3$ that meet predefined confidence thresholds or validation criteria are retained for synthesis. Retrieval-Augmented Generation (RAG) techniques may be optionally employed here to provide the LLM with access to up-to-date vulnerability databases or best practice guidelines during both assessment and validation, further enhancing accuracy.

Validated findings from the execution stage ($s_3$) are aggregated and synthesized to build a coherent and prioritized overview of the contract's security posture ($s_4$). This involves more than just listing vulnerabilities. It assigns severity levels to each finding based on factors like potential business impact (e.g., financial loss, loss of control, denial of service) and exploitability (e.g., conditions required, attacker privileges needed, public knowledge of exploit). The synthesis process also aims to identify relationships between vulnerabilities, such as shared root causes or dependencies. This systematic classification and correlation allow the system to filter and present findings in a manner that highlights the most urgent issues requiring attention.

The final stage focuses on producing a clear, concise, and actionable audit report. This report consolidates the initial contract understanding ($s_1$), the audit plan ($s_2$), and the synthesized findings ($s_4$), including severity assessments and, where possible, actionable recommendations for mitigation. The structure and language of the report can be tailored to suit different audiences (e.g., developers requiring technical details, or management needing high-level summaries). Drawing insights from best practices in security reporting and relevant studies \cite{liu2024exploring, yin2024multitask}, the report aims to cover key dimensions such as clear vulnerability descriptions, severity evaluations, precise localization of issues in the code, and concise summaries. 

\subsubsection{Task-Specific Review and Calibration}
Given that LLMs can occasionally "hallucinate" or generate outputs that deviate from factual contract code \cite{bender2021dangers}. Stage A3 is explicitly structured into two critical steps for each sub-task: (1) initial task-specific review and (2) rigorous calibration. This two-step process is designed to maximize accuracy and minimize false positives or missed vulnerabilities before findings are aggregated.

\paragraph{(1) Task-Specific Review ($e^i$)}
In this initial step, the LLM is prompted to investigate one narrowly defined security concern as dictated by the sub-task $t^i$ from the audit plan. The objective is to have the LLM scrutinize specific code segments or contract behaviors relevant to that concern. Example prompts for this stage might include:

\begin{itemize}
  \item “\textit{Analyze function \texttt{withdrawFunds()} for potential reentrancy vulnerabilities. Detail any observed patterns and the conditions under which they might be exploitable.}”
  \item “\textit{Examine all division operations within this contract for susceptibility to integer overflow or underflow issues. List each susceptible operation and its location.}”
\end{itemize}

By focusing on a singular security issue, the LLM's analysis remains targeted, leveraging its understanding of contract logic and known security best practices to pinpoint potential anomalies. The output of this step for each sub-task $t^i$ is an initial review finding $e^i$, formally expressed as:

\begin{equation}
  e^i = \mathcal{M}(\mathcal{P}_e^i(t^i, c))
\end{equation}
where $\mathcal{P}_e^i$ is the task-specific review instruction for sub-task $t^i$, $c$ is the relevant smart contract code (or snippets thereof), and $\mathcal{M}$ represents the LLM.

\paragraph{(2) Calibration $v^i$}
Following the initial review, the calibration step aims to validate, refine, and expand the preliminary findings $e^i$. This involves guiding the LLM to critically re-examine its initial assessment. The LLM is challenged to ground its claims firmly in the contract's code and relevant security principles. An example calibration prompt could be:

“\textit{Explain how the identified vulnerability in $e^i$ can be exploited based on the contract’s actual variables and functions. Provide a code snippet highlighting the relevant lines. Also describe any defenses that might already exist in the code.}”

This self-correction and evidence-gathering process helps to filter out unsupported assertions and reduce LLM hallucinations. The output is a calibrated finding $v^i$:
\begin{equation}
    v^i = \mathcal{M}(\mathcal{P}_v^i(e^i, t^i))
    \label{eq:calibration}
\end{equation}
where $\mathcal{P}_v^i$ is the calibration instruction, which uses the initial review $e^i$, the original sub-task context $t^i$.

\subsubsection{RAG-Enhanced Calibration for Improved Accuracy and Grounding}
\label{subsubsec:rag_enhancement}
While pre-trained LLMs possess extensive general knowledge, their understanding of domain-specific knowledge can be limited or outdated. This can lead to gaps in analysis or less precise explanations, particularly for edge-case scenarios. To mitigate this, we enhance the calibration step with Retrieval-Augmented Generation (RAG) \cite{lewis2020retrieval}, as illustrated in Figure~\ref{fig:rag}. RAG augments the LLM's internal knowledge by providing access to relevant excerpts from external, up-to-date knowledge sources during generation.

We have observed instances where an LLM might correctly identify a vulnerable code location but provide an incomplete or imprecise explanation, especially for complex or uncommon vulnerabilities. RAG addresses this by retrieving documents that offer detailed, authoritative explanations for similar issues. Within our Execute Phase, each calibration prompt $\mathcal{P}_v^i$ is augmented with information retrieved from a curated knowledge base. This knowledge base includes smart contract databases (e.g., SWC Registry, CWE, CVE), security guidelines (e.g., Ethereum Foundation, ConsenSys Diligence), seminal research papers, technical blogs detailing novel exploits, and community-vetted best practice repositories (see Appendix \ref{sec:rag_baseknowledge} for a list of sources). These resources are embedded and stored in a Vector Database (using models like \textit{Sentence-BERT}~\cite{reimers2019sentence}) for efficient semantic retrieval.

The RAG-enhanced calibration workflow proceeds as follows:
\begin{enumerate}
  \item Formulate Retrieval Query ($Q_v^i$): Based on the sub-task $t^i$ and the initial review output $e^i$, a targeted retrieval query is formulated. This query typically includes keywords related to the potential vulnerability (e.g., “integer division,” “reentrancy,” “\texttt{call.value} usage”), function names, and other contextual cues from $t^i$ and $e^i$.
  \item Retrieve Relevant Documents ($D_v^i$): The system queries the VectorDB using $Q_v^i$ to retrieve the top-$k$ most relevant document snippets $D_v^i = \{d_1, d_2, \dots, d_k\}$.
  \item Evidence-Grounded Calibration: The LLM then synthesizes the information from these retrieved documents $D_v^i$ with the initial review output $e^i$ and the original sub-task context $t^i$ to produce the calibrated finding $v^i$. The calibration instruction $\mathcal{P}_v^i$ is structured to explicitly instruct the LLM to incorporate and reference the retrieved evidence. This is formally represented as:
  \begin{equation}
  v^i = \mathcal{M}\bigl(\mathcal{P}_v^i(e^i, t^i, c, D_v^i)\bigr), \text{ where } D_v^i = \mathrm{Retrieve}(Q_v^i)
  \label{eq:rag_calibration}
  \end{equation}
\end{enumerate}

By compelling the LLM to ground its calibration in authoritative, domain-specific references, we significantly enhance the accuracy, reliability, and explanatory depth of the findings. Each calibrated finding $v^i$ that emerges from this RAG-enhanced process is thus considered a robust and well-substantiated observation, ready for aggregation in Stage A4. Additional details, including further instances and sources for RAG design, are provided in Appendix~\ref{sec:rag_baseknowledge}

The final output of the Execute Phase (Stage A3) is a curated list of \emph{validated findings} ($v^i$), each substantiated by evidence from the contract code and, where applicable, external knowledge sources via RAG. These high-confidence findings are then passed to Stage A4 (Findings Synthesis), where they are further analyzed, prioritized by severity, and checked for inter-dependencies.

\subsection{LLM-Powered Audit Evaluator}
\label{subsec:llm_evaluator}
Assessing the quality of the generation is an even more arduous task than the generation itself, and this issue has not been given adequate consideration. Few studies have pointed that LLM has been a good evaluator for natural language generation (NLG), and the existing approaches are not tailored to the specific needs of the audit process \cite{fu2023gptscore, liu2023g}. 
Inspired by RAGAs \cite{es2024ragas}, we designed and implemented an automated LLM-Powered Audit Evaluator. 
This evaluator systematically compare the findings generated by {SmartAuditFlow} against expert-annotated ground truth answers from the same expert audited-report.
The primary objectives of this evaluator are to ensure fair and reproducible assessments, accelerate the evaluation cycle, and provide actionable insights for the continual improvement of the {SmartAuditFlow} framework. Our LLM-Powered Audit Evaluator operates in two main phases: (1) LLM-Driven Finding Comparison and (2) Quantitative Performance Scoring.

The evaluator requires two structured inputs for each smart contract assessed: (i) the \textbf{{SmartAuditFlow}-generated report}, detailing each detected vulnerability with its type, natural language description, specific code location(s), and assigned severity; and (ii) the corresponding \textbf{Standard Answer report}. Standard Answers are meticulously curated from existing human-expert audit reports for the identical smart contract version, transformed into the same structured format as the {SmartAuditFlow} output to facilitate direct comparison. This ground truth includes expert-validated vulnerability types, descriptions, locations, and severities.

\paragraph{LLM-Driven Finding Comparison} This phase focuses on accurately mapping each finding reported by {SmartAuditFlow} to its corresponding entry in the Standard Answer, or identifying it as unique. An LLM is central to this process, leveraging its semantic understanding capabilities to interpret nuanced textual descriptions and contextual information. The key steps include:

\begin{itemize}
  \item \textbf{Finding Pairing:} For each vulnerability reported by {SmartAuditFlow}, the LLM identifies the most plausible candidate match(es) from the Standard Answer. This step also considers findings in the Standard Answer to identify those potentially missed by {SmartAuditFlow}.
  \item \textbf{Multi-Dimensional Semantic Assessment:} Each candidate pair (one finding from {SmartAuditFlow}, one from the Standard Answer) is subjected to a detailed comparative analysis by an LLM. This assessment uses a predefined rubric considering multiple dimensions: (a) vulnerability type; (b) {description semantic similarity} (assessing if both descriptions refer to the same underlying issue); and (c) {code location precision} (e.g., exact line match, function-level match, overlapping blocks).
  \item \textbf{Match Categorization:} Based on the multi-dimensional assessment and the rubric, the LLM categorizes the relationship between the paired findings. Inspired by prior research \cite{wei2024measuring}, we adopt a nuanced classification approach rather than a binary true/false distinction. The primary categories include \texttt{Exact Match} (all critical aspects align), \texttt{Partial Match} (core issue identified but differs on a specific dimension like severity or location precision), and \texttt{Incorrect} (the response misidentifies the core issue in the Standard Answer).
\end{itemize}

\paragraph{Quantitative Performance Scoring} This phase translates the categorized outputs from the comparison phase into quantitative metrics to evaluate {SmartAuditFlow}'s overall performance. For metrics like accuracy, MRR and MAP, findings in the {SmartAuditFlow} report are assumed to be ranked (e.g., by severity or an internal confidence score). These metrics are described in Section~\ref{subsubsec:evaluation_criteria}.

\section{Evaluation}
\label{sec:experiments}
This section presents the evaluation of our proposed framework.
We assess its performance in detecting smart contract vulnerabilities and compare it against established tools and other LLM-based methods.

\subsection{Research Questions}
Our evaluation is guided by the following research questions (RQs):
\begin{itemize}
  \item \textbf{RQ1: How effectively does SmartAuditFlow identify common vulnerability types?} This question investigates the models’ ability to detect known vulnerabilities as defined in the SWC-registry. We compare the performance of our framework against other methods in identifying common vulnerabilities. 
  \item \textbf{RQ2: How does {SmartAuditFlow} perform when evaluated using different top-$N$ thresholds?}
  We assess the performance of {SmartAuditFlow} by considering vulnerabilities identified within the top-$N$ ranked results (e.g., N=1, 5, max). This investigates the framework's ability to prioritize critical issues effectively.
  \item \textbf{RQ3: How does SmartAuditFlow perform on complex real-world projects?} Real-world smart contracts often exhibit greater complexity than curated benchmark datasets. We evaluate {SmartAuditFlow} on a dataset of audited real-world projects and compare its performance to existing tools to assess its practical applicability and scalability.
  \item \textbf{RQ4 What is the impact of different underlying LLMs on {SmartAuditFlow}'s performance?} With the rapid evolution of LLMs, we investigate how the choice of the LLM backbone (e.g., models from the GPT series, Claude series, DeepSeek, Gemini family) affects {SmartAuditFlow}'s vulnerability detection accuracy and overall effectiveness.
  \item \textbf{RQ5: How does the integration of different external knowledge sources influence {SmartAuditFlow}'s performance?}  We conduct an ablation study to assess the impact of integrating static analysis tool outputs and RAG on the overall performance of {SmartAuditFlow}, compared to a baseline version without these enhancements.
  \item \textbf{RQ6: How does {SmartAuditFlow} compare to other contemporary LLM-based vulnerability detection methods?} We evaluate our method on a relevant dataset (the CVE set), comparing its ability to correctly identify vulnerabilities and generate comprehensive, accurate descriptions against other LLM-based approaches.
\end{itemize}

\subsection{Experimental Setup}

\subsubsection{Dataset}
\label{subsubsec:datasets}
To rigorously evaluate SmartAuditFlow, we utilize a diverse Set of datasets that reflect various scenarios and complexities in smart contract auditing. These include:

\begin{itemize}
  \item \textbf{Prompt Optimization Set}: This dataset comprises 1,000 smart contracts from various domains (e.g., DeFi, NFTs, gaming), selected to encompass a wide range of contract structures and potential security concerns. This set is exclusively used for the iterative optimization of prompt instructions ($\rho$) for Stages A1 and A2 within our Plan Phase, as detailed in Section \ref{subsec:planning_phase_details}. Statistics and construction details for this dataset are provided in Appendix A (Table \ref{tab:dataset_statistics}).
  \item \textbf{Standard Vulnerability Set}: This dataset contains smart contracts annotated with common vulnerability types, which serves as a benchmark for evaluating the detection of well-understood vulnerability types against established tools. We utilize the SmartBugs-curated dataset \cite{durieux2020empirical}, which is widely adopted by both developers and researchers. The dataset includes 143 contracts, each annotated with common vulnerabilities based on the DASP classification, making it an excellent resource for benchmarking vulnerability detection tools.
  \item \textbf{Real-world Contracts Set}: To assess performance on more complex, real-world scenarios, we use a curated subset of contracts from projects audited by Code4rena \cite{durieux2020empirical}. This set consists of 72 projects, encompassing 6,454 contracts, among which 243 contracts have known issues, including 784 instances of high or medium-severity vulnerabilities. This dataset tests {SmartAuditFlow}'s capability to handle complex contract logic and diverse vulnerability manifestations.
  \item \textbf{CVE Set}:  We curated a set of Common Vulnerabilities and Exposures (CVEs) relevant to smart contracts. As of early 2025, there are 592 smart contract-related CVEs, with integer overflows being a predominant category.Followed by another co-worker's PropertyGPT \cite{liu2024propertygpt},  we selected 13 representative CVEs for detailed analysis, including three integer overflow cases, three access control vulnerabilities, and a diverse set of four other logic bugs.
\end{itemize}

All curated datasets, including our annotations, ground truth mappings, and processing scripts, are made openly accessible via our GitHub repository to foster reproducibility and community-driven extensions.

We also explored other publicly available datasets, such as DeFiVulnLabs \footnote{https://github.com/SunWeb3Sec/DeFiVulnLabs} and DeFiHackLabs \footnote{https://github.com/SunWeb3Sec/DeFiHackLabs}. For {DeFiVulnLabs} dataset , the GPT-4o model achieving a score of 37/50. DeFiHackLabs, primarily consisting of proof-of-concept exploit pieces of code snippets, was deemed more suitable for attack replication studies than for systematic vulnerability detection benchmarking of full contracts. As a result, we did not include these two datasets in our paper.

\subsubsection{Evaluation Criteria}
\label{subsubsec:evaluation_criteria}
The primary objective of our evaluation is to assess the effectiveness of the SmartAuditFlow framework in identifying vulnerabilities in smart contracts. This involves a systematic comparison of {SmartAuditFlow}'s findings against the ground truth for each contract. The classification outcomes are defined as:

\begin{itemize}
  \item \textbf{True Positive (TP)}: The framework correctly identifies a vulnerability that exists in the contract, as validated against the ground truth through our LLM-Powered Audit Evaluator.
  \item \textbf{False Positive (FP)}: The framework fails to detect a vulnerability that is present in the contract according to the ground truth.
\end{itemize}

For evaluating the ranking quality of vulnerabilities reported by the framework:
\paragraph{Top-$N$ Accuracy}
We assess if a true vulnerability is identified within the top-$N$ findings reported by the framework (e.g., for N=1, 3, 5). This method is useful, especially in complex tasks like smart contract vulnerability detection, where vulnerabilities may not always be ranked perfectly.

\paragraph{Mean Reciprocal Rank (MRR)}
This metric evaluates how highly the first correctly identified relevant vulnerability is ranked for each contract (or query). A higher MRR indicates that the method tends to rank true vulnerabilities closer to the top. The formula is:

\begin{equation}
  MRR = \frac{1}{|Q|} \sum_{i=1}^{|Q|} \frac{1}{rank_i}
\end{equation}
where $|Q|$ is the total number of queries, and $\text{rank}_i$ is the rank of the first TP for the $i$-th query (or 0 if no TP is found).

\paragraph{Mean Average Precision (MAP)}
MAP provides a comprehensive measure of both precision and the ranking order of all identified vulnerabilities. It is the mean of Average Precision (AP) scores across all contracts/queries. AP for a single query is calculated as:
\begin{equation}
  AP = \frac{1}{N} \sum_{n=1}^{N} P(n) \cdot \text{rel}(n)
\end{equation}
where \( P(n) \) is the precision at rank \(n\) (i.e., the fraction of TPs in the top \(n\) results), and \( \text{rel}(n) \) is the relevance of the \(n\)-th result, where 1 indicates a TP and 0 indicates a FP.

\subsection{Implementation Details}
All experiments for {SmartAuditFlow} and static tools were conducted within a Docker container to ensure a consistent, isolated, and reproducible environment across all test runs.
For evaluating the vulnerability detection capabilities involving LLMs, we adopted a single attempt approach (often denoted as pass@1) \cite{wei2024measuring}. This reflects practical auditing scenarios where auditors typically seek an effective assessment in one primary attempt. 

As detailed in Section \ref{subsec:planning_phase_details}, our framework employs an iterative strategy to optimize prompt instructions ($\rho$) for Stages A1 and A2. For the initial seeding of the evolutionary prompt optimization process (i.e., generating the diverse starting population of candidate prompts), we utilized GPT-4o (model versiongpt-4o-2024-11-20) as an "assistant LLM" to generate an initial set of 20 diverse prompt candidates for both A1 and A2 tasks. The Prompt Optimization Set (described in Section \ref{subsubsec:datasets}) was partitioned into a 70\% subset for running the evolutionary optimization algorithm and a 30\% subset for validating the performance of the finally selected optimized prompts. Illustrative examples of both initial and optimized prompts are provided in Appendix A.

For the majority of our experiments evaluating the core capabilities and comparative performance of {SmartAuditFlow}, we employed GPT-4o (gpt-4o-2024-11-20) s the default LLM integrated within the framework. This model was selected based on its strong performance demonstrated in our preliminary assessments and its advanced reasoning capabilities. We systematically substituted the primary operating LLM within the framework with several other leading models: Gemini-2.5-pro (gemini-2.5-pro-preview-03-25), Claude-3.7-sonnet (claude-3-7-sonnet-20250219), and DeepSeek-v3.  These models were accessed via their respective public APIs.

\newmdenv[
  backgroundcolor=gray!10, 
  linewidth=0.5pt, 
  roundcorner=10pt,
  skipabove=\baselineskip
]{custommdframed}

\subsection{Experimental Results}
To answer the research questions posed earlier, we present the experimental results.

\begin{figure}
  \centering
  \includegraphics[width=0.7\textwidth]{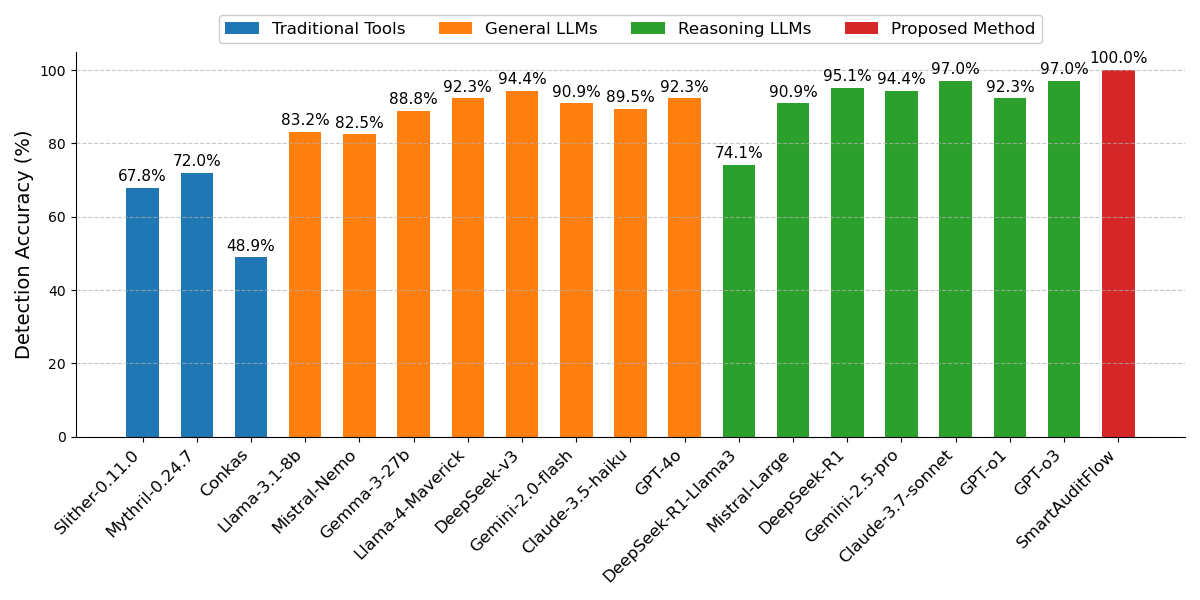}
  \caption{Evaluation of Smart Contract Vulnerability Detection Tools - A Comparative Analysis}
  \label{fig:comparative_rq1}
\end{figure}

\subsubsection{RQ1: Effectiveness in Identifying Common Vulnerabilities}
\label{subsubsec:rq1_effectiveness}
To address RQ1, we evaluated {SmartAuditFlow}'s effectiveness in identifying common vulnerability types using the Standard Vulnerability Set. We benchmarked {SmartAuditFlow} against a comprehensive suite of baselines, including established traditional static analysis tools and a range of standalone LLMs prompted with a standardized, direct querying strategy.
The traditional static analysis tools selected for comparison were Slither (0.11.0), Mythril (0.24.7), and Conkas. For standalone LLM baselines, we categorized and evaluated two types of models:
\textbf{General-Purpose Instruction-Tuned Models}: Llama-3.1-8b-it, Mistral-Nemo-Instruct, Gemma-3-27b-Instruct, Llama-4-Maverick-Instruct, DeepSeek-v3,  Gemini-2.0-flash, Claude-3.5-haiku (claude-3-5-haiku-20241022), and GPT-4o (gpt-4o-2024-11-20);
\textbf{Reasoning Models}: DeepSeek-R1-Llama3-8b, Mistral-Large (mistral-large-latest), DeepSeek-R1, Gemini-2.5-pro (gemini-2.5-pro-preview-03-25), Claude-3.7-sonnet (claude-3-7-sonnet-20250219), GPT-o1 (o1-2024-12-17), and GPT-o3 (o3-2025-04-16).

For evaluating the detection capability of all methods, including standalone LLMs, we considered a true vulnerability to be detected if it was correctly identified and reported anywhere in the method's output (effectively a recall-oriented measure for each ground truth instance). The results of this comparative evaluation are summarized in Figure \ref{fig:comparative_rq1}.

Our findings indicate that {SmartAuditFlow} (utilizing its primary configuration with GPT-4o, unless specified otherwise) achieves a detection accuracy of 100\% on this dataset. This performance significantly surpasses that of the traditional static analysis tools. For instance, the best-performing traditional tool, Mythril, achieved an accuracy of 72.0\%. Notably, even the standalone LLM baselines generally outperformed traditional tools; for example, the lowest-performing LLM in our set, DeepSeek-R1-Llama3-8B, still achieved 74.1\% accuracy. This underscores the substantial potential of LLMs in vulnerability detection.

Furthermore, {SmartAuditFlow} also demonstrated a clear advantage over standalone LLM applications. As an example of its workflow-driven enhancement, when {SmartAuditFlow} was configured to use GPT-4o as its internal LLM, it achieved the 100\% accuracy, representing a 7.7 percentage point improvement over using GPT-4o (pass@1). 
The results also reveal that reasoning models exhibited superior performance compared to their general-purpose models from the same family or size class. For instance, GPT-o3 (97.0\%) and Claude-3.7-Sonnet (97.0\%) outperformed models like the baseline GPT-4o (92.3\%) and Claude-3.5-Haiku (89.5\%). The results also confirmed the general trend that larger-parameter models and those specifically optimized for reasoning or coding tasks in this domain compared to smaller or more generic instruction-tuned models.

\begin{custommdframed}
  \textbf{Answer to RQ1:} SmartAuditFlow demonstrates high effectiveness in identifying common smart contract vulnerabilities, achieving a detection accuracy of 100\% on the Standard Vulnerability Set. This significantly outperforms both traditional static analysis tools and standalone applications of advanced LLMs. The results affirm that strategically applying LLMs within a structured workflow like {SmartAuditFlow} substantially enhances their intrinsic capabilities for vulnerability detection.
\end{custommdframed}

\subsubsection{RQ2: Performance Across Different Top-$N$ Thresholds and Ranking Quality}
\label{subsubsec:rq2_top_n}
While RQ1 assessed overall detection capabilities at top-$max$ accuracy, practical auditing often involves reviewing a limited subset of reported findings. Auditors may prioritize inspecting only the top few (e.g., top-1, top-5, or top-10) predictions, especially when dealing with tools that generate a large number of outputs, which can include false positives or less critical information.
Therefore, for RQ2, we evaluate how effectively different methods rank true vulnerabilities by examining their performance at various top-$N$ thresholds (top-1, top-5, and top-$max$). We also measure the MRR and report the average number of findings generated per contract by each method.

Table \ref{tab:model_performance} demonstrates superior performance across all top-$N$ accuracy metrics, achieving \textbf{66.4\% at top-1}, \textbf{99.3\% at top-5}, and \textbf{99.2\% at top-$max$}. This indicates that not only does {SmartAuditFlow} detect a comprehensive set of vulnerabilities, but it also ranks them highly. For comparison, a high-performing standalone LLM baseline, GPT-o3, achieved 58.7\% at top-1, 93.0\% at top-5, and 97.0\% at top-$max$. As expected, for all models, detection accuracy increases with larger $N$, underscoring the utility of ranked outputs; however, the rate of improvement and the accuracy achieved at lower $N$ values are key differentiators.

The MRR scores further validate {SmartAuditFlow}’s superior ranking capabilities. With an \textbf{MRR of 0.8}, {SmartAuditFlow} ranks the first true positive vulnerability, on average, at approximately position 1.25 (calculated as $1/0.8 = 1.25$). This implies that auditors using {SmartAuditFlow} would typically encounter the first critical issue within the top one or two reported items. In contrast, other strong standalone LLMs like Claude-3.7-sonnet (with an MRR of 0.73) and GPT-o3 (MRR of 0.72) rank the first true positive at average positions of approximately 1.37 and 1.39, respectively. Models with lower MRR scores, such as DeepSeek-R1-Llama3-8B (MRR of 0.54, corresponding to a top-5 accuracy of 73.4\%), require auditors to sift through more findings to locate the first true positive.

\begin{figure}
  \centering
  \includegraphics[width=0.72\textwidth]{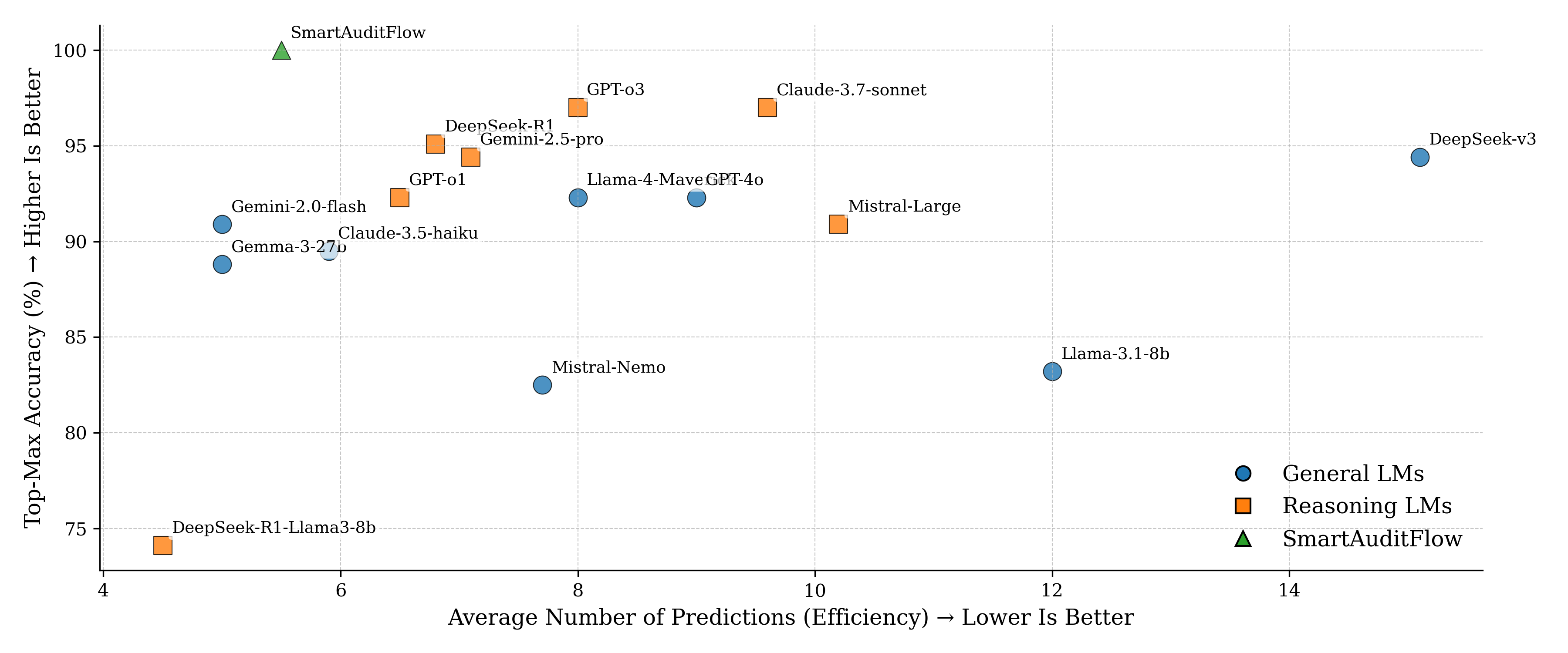}
  \caption{Trade-off Between Accuracy (top-max) and Efficiency (Average Number of Predictions)}
  \label{fig:tradeoff_top_accuracy}
\end{figure}

Efficiency, in terms of the average number of findings reported per contract, is also a critical factor. Figure \ref{fig:tradeoff_top_accuracy} illustrates the trade-off between top-$max$ accuracy and this output volume. Some models, like DeepSeek-v3 (achieving 94.4\% top-$max$ accuracy), generate a high number of average outputs (e.g., 15.1 per contract),  which improves overall recall (top-$max$) but potentially increases auditor workload due to a higher number of findings to review (including potential false positives). General-purpose instruction-tuned models such as Llama-3.1-8b-it and Mistral-Nemo-Instruct achieved moderate top-$max$ accuracy (around 83\%) but also produced a relatively high average of 7.7 to 12 findings. Reasoning-enhanced standalone LLMs (e.g., DeepSeek-R1, Claude-3.7-sonnet, and GPT-o3) demonstrated improved top-$max$ accuracy (94-97\%) with a more moderate average of 6.5 to 9.6 findings. \texttt{SmartAuditFlow} stands out by achieving the highest top-$max$ accuracy (100\%) while generating an average of only \textbf{5.5 findings per contract}.

\begin{custommdframed}
  \textbf{Answer to RQ2:} {SmartAuditFlow} excels in ranking critical vulnerabilities, achieving a top-$max$ accuracy of 100\%. Its high MRR of 0.8 indicates that the first true positive is typically found very early in the ranked list (average rank ~1.25). Furthermore, {SmartAuditFlow} demonstrates superior efficiency by achieving this high level of detection accuracy while generating a significantly lower average number of findings per contract (5.5) compared to other standalone LLM approaches. 
\end{custommdframed}

\begin{table*}[t]
  \centering
  \caption{Evaluation of Detection Ability across Different Models - A Comparative Analysis}
  \small
  \label{tab:model_performance}
  \begin{threeparttable}
  \begin{tabular}{l | ccc|c| c}
    \toprule
    \textbf{Tool} & \textbf{top-1} & \textbf{top-5} & \textbf{top-max} & \textbf{MRR} & \textbf{Avg. Outputs}  \\
    \midrule
    Llama-3.1-8b-Instruct & 37.1\%  & 79.0\% & 83.2\% & 0.54 & 12.0 \\ 
    Mistral-Nemo-Instruct & 37.1\% & 76.9\% & 82.5\% & 0.51 & 7.7  \\
    Gemma-3-27b-it & 51.0\% & 87.4\% & 88.8\%  & 0.65 & 5.0  \\
    Llama-4-Maverick-Instruct & 38.5\% & 84.6\% & 92.3\% &  0.55 & 8.0 \\ 
    DeepSeek-v3 & 47.6\% & 79.0\% & 94.4\% & 0.60 & 15.1 \\ 
    Gemini-2.0-flash & 62.9\% & 87.4\% & 90.9\% & 0.72 & 5.0 \\ 
    Claude-3.5-haiku & 60.1\% & 88.1\% & 89.5\% & 0.70 & 5.9 \\ 
    GPT-4o & 49.0\% & 86.7\% & 92.3\%  &0.65 & 9.0 \\ \midrule
    DeepSeek-R1-Llama3-8b & 42.0\% & 73.4\% & 74.1\%  & 0.54 & 4.5  \\
    Mistral-Large & 18.2\% & 84.6\% & 90.9\%  & 0.46 & 10.2   \\
    DeepSeek-R1 & 56.6\% & 93.0\% & 95.1\% & 0.69 &6.8\\
    Gemini-2.5-pro & 58.0\% & 90.9\% & 94.4\% & 0.71 & 7.1 \\ 
    Claude-3.7-sonnet & 60.8\% & 93.0\% & 97\% & 0.73 & 9.6 \\ 
    GPT-o1 & 54.5\% & 88.8\% & 92.3\% & 0.67 & 6.5\\ 
    GPT-o3 & 58.7\% & 93.0\% & 97.0\% & 0.72 & 8.0\\
    \midrule
    SmartAuditFlow & 66.4\% & 99.3\% & 100\%  & 0.80 & 5.5\\
    \bottomrule
  \end{tabular}
  \end{threeparttable}
\end{table*}

\subsubsection{RQ3: Performance on Real-World Smart Contracts}
\label{subsubsec:rq3_real_world}
While RQ1 and RQ2 demonstrated strong performance on datasets with common, often well-defined vulnerability types, real-world smart contracts frequently present more complex, nuanced, and unique security issues. To address RQ3, we conducted a comprehensive evaluation of {SmartAuditFlow} and selected baseline methods using our Real-World Contracts Set. This evaluation primarily utilizes the top-$max$ accuracy approach to assess overall detection capability on these challenging contracts.

Based on findings from previous experiments (RQ1, RQ2) and their prevalence, we selected a representative set of standalone LLMs for comparison: General-Purpose models: DeepSeek-v3, Gemini-2.0-flash, Claude-3.5-haiku, GPT-4o; Reasoning models: DeepSeek-R1, Gemini-2.5-pro, Claude-3.7-sonnet, GPT-o3. We also evaluated traditional static analysis tools, Slither and Mythril, on this Real-World Contracts Set. Consistent with findings in related work \cite{durieux2020empirical, sun2024gptscan}, these tools exhibited very limited capability in detecting the complex logic vulnerabilities, often failing to identify any of the nuanced issues present.

The performance results on the Real-World Contracts Set are detailed in Table \ref{tab:realworld_performance}. {SmartAuditFlow} achieved a top-$max$ accuracy of \textbf{32.4\%} and an MAP of \textbf{0.334}, while generating the lowest average number of outputs per contract at \textbf{6.2}. This relatively lower accuracy compared to the Standard Vulnerability Set underscores the significantly increased difficulty of real-world contracts. Importantly, this performance represents an approximate 11.2 percentage point improvement over a baseline application of GPT-4o using pass@1. 

The MAP of 0.334 for {SmartAuditFlow} indicates that, even on these complex contracts, the first true positive vulnerability is typically found relatively high in its ranked list of findings (average rank approx. $1/0.334 \approx 2.99$). The low average output count (6.2) further suggests that {SmartAuditFlow} is comparatively precise, minimizing the review burden on developers.

\begin{figure}[h]
  \centering
  \includegraphics[width=0.6\textwidth]{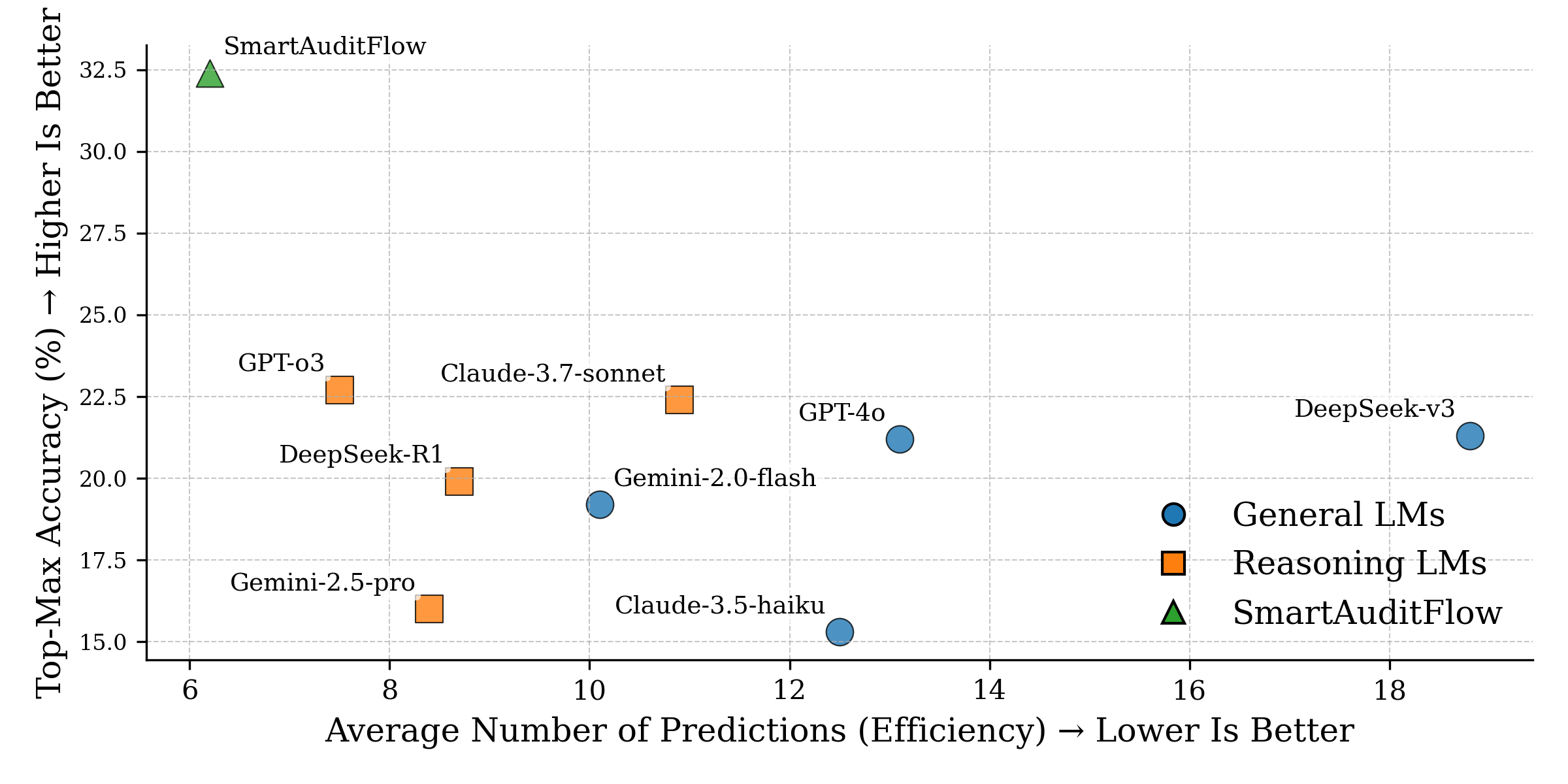}
  \caption{Trade-off Between Accuracy (top-max) and Efficiency (Average Number of Predictions)}
  \label{fig:realworld_performance}
\end{figure}

Figure \ref{fig:realworld_performance} illustrates the trade-off between top-$max$ accuracy and prediction efficiency (average number of outputs per contract). On this challenging dataset, many standalone LLMs, whether general-purpose or reasoning, struggled to achieve high accuracy without generating a large number of outputs.
For example, general-purpose models like DeepSeek-v3 and GPT-4o, while achieving top-$max$ accuracies in the range of 15\% to 21.3\%, produced a higher average of 13 to 18.8 findings. Reasoning models achieved accuracies in the range of 16-23\% with a somewhat more moderate 7.5 to 11 findings on average.
In stark contrast, {SmartAuditFlow} achieved the highest recorded top-$max$ accuracy of 32.4\% with exceptional efficiency, averaging only 6.2 findings.

\begin{custommdframed}
  \textbf{Answer to RQ3:} On the challenging Real-World Contracts Set, {SmartAuditFlow} demonstrates significantly better performance than standalone LLM applications and traditional tools. It achieves a top-$max$ detection accuracy of 32.4\% and an MAP of 0.334, while maintaining high efficiency by generating only an average of 6.2 findings per contract. This underscores {SmartAuditFlow}'s enhanced capability to accurately and efficiently identify vulnerabilities in complex, real-world smart contract scenarios.
\end{custommdframed}

\begin{table*}[t]
  \centering
  \caption{Evaluation of Smart Contract Vulnerability Detection Tools: A Comparative Analysis}
  \small
  \label{tab:realworld_performance}
  \begin{threeparttable}
  \begin{tabular}{l | c | c | c}
    \toprule
    \textbf{Tool} & \textbf{Accuracy (top-max)} & \textbf{Avg. Outputs} & \textbf{MAP}\\
    \midrule
    DeepSeek-v3 & 21.3\% & 18.8 & 0.148 \\
    Gemini-2.0-flash & 19.2\% & 10.1 & 0.203\\ 
    Claude-3.5-haiku & 15.3\%  & 12.5 & 0.123\\
    GPT-4o & 21.2\%  & 13.1 & 0.170 \\ \midrule
    DeepSeek-R1 & 19.9\% & 8.7 & 0.267 \\
    Gemini-2.5-pro  & 16.0\% & 8.4 & 0.199 \\
    Claude-3.7-sonnet & 22.4\% & 10.9 & 0.211 \\
    GPT-o3 & 22.7\% & 7.5 & 0.236 \\ \midrule
    SmartAuditFlow & 32.4\% & 6.2 & 0.334 \\
    \bottomrule
  \end{tabular}
  \end{threeparttable}
\end{table*}

\subsubsection{RQ4: Impact of Different LLM Backbones on {SmartAuditFlow} Performance}
\label{subsubsec:rq4_configurations}

{SmartAuditFlow} is designed as a versatile framework whose performance can be influenced by the choice of the underlying LLM and other operational parameters. To address RQ4, we specifically investigate the impact of varying the LLM backbone on {SmartAuditFlow}'s vulnerability detection accuracy and operational efficiency. For this analysis, all other parameters of the {SmartAuditFlow} workflow, especially the optimized prompt sets, were held constant to isolate the effect of the LLM choice. The evaluations for RQ4 were conducted on the Real-World Contracts Set to assess performance on complex, practical scenarios.

We configured {SmartAuditFlow} with four distinct high-performance LLMs: GPT-4o, DeepSeek-v3, Claude-3.7-sonnet, and Gemini-2.5-pro. Performance was compared across four key metrics: top-$max$ accuracy, MAP for ranking quality, average number of findings reported per contract, and average inference steps required by \texttt{SmartAuditFlow} per contract.

The results in Table \ref{tab:configurations_performance} reveal that \texttt{SmartAuditFlow}'s performance, while generally strong, varies meaningfully with the integrated LLM. \texttt{SmartAuditFlow} configured with Gemini-2.5-pro attained the highest top-$max$ accuracy at \textbf{37.16\%}, closely followed by Claude-3.7-sonnet configuration at \textbf{36.82\%}. However, the Gemini configuration also produced the highest average number of findings per contract (16.9), suggesting that its high recall might come with increased output verbosity, potentially requiring more auditor review time. In contrast, the GPT-4o configuration, while achieving a top-$max$ accuracy of 32.4\%, excelled in ranking quality, securing the highest MAP of \textbf{0.334}.

The Claude-3.7-sonnet configuration presented a compelling balance: achieving nearly top-tier accuracy (36.82\%), strong ranking quality (MAP of 0.315), and a moderate number of average outputs 12.8. However, this configuration registered a higher average number of inference steps 34.2 compared to others like  Gemini (30.1), suggesting potentially higher computational cost or latency within our workflow. The DeepSeek-v3 configuration, conversely, exhibited lower performance across most metrics in this demanding real-world context when integrated into {SmartAuditFlow}, indicating it might be less suited for such complex vulnerability detection tasks within our framework compared to the other evaluated LLMs.

\begin{custommdframed}
  \textbf{Answer to RQ4:} {SmartAuditFlow}'s performance is notably influenced by the choice of its LLM backbone, allowing for optimization based on specific auditing priorities. Gemini-2.5-pro can achieve the highest top-$max$ detection accuracy (37.16\%) but may produce more verbose outputs. GPT-4o can provide superior ranking of the most critical vulnerabilities and potentially more concise reports. Claude-3.7-sonnet offers a strong balance between high detection accuracy and good ranking quality.
\end{custommdframed}

\begin{table*}[t]
  \centering
  \caption{Evaluation on Different LLM for SmartAuditFlow}
  \small
  \label{tab:configurations_performance}
  \begin{threeparttable}
  \begin{tabular}{l | c | c | c | c }
    \toprule
    \textbf{Configuration LM} & \textbf{Avg. (top-max)} & \textbf{Avg. Outputs} & \textbf{MAP} & \textbf{Avg. Steps}  \\
    \midrule
    GPT-4o & 32.4\% & 6.2 & 0.334 & 32.1\\
    DeepSeek-v3 & 25.1\% & 9.6 & 0.174 & 38.7  \\
    Claude-3.7-sonnet & 36.8\% & 12.8 & 0.315 & 34.2\\
    Gemini-2.5-pro & 37.2\% & 16.9 & 0.253 & 30.1 \\
    \bottomrule
  \end{tabular}
  \end{threeparttable}
\end{table*}

\subsubsection{RQ5: Impact of Incorporating External Knowledge}
\label{subsubsec:rq5_external_knowledge}
To address RQ5, we investigated the impact of incorporating external knowledge on the performance of SmartAuditFlow by comparing its accuracy and efficiency with and without external knowledge, including the combination of static tools (see Section 3.2.3) and RAG techniques (see Section 3.3.2). As established in RQ4 results, the baseline configuration of {SmartAuditFlow} utilized Gemini-2.5-pro as its LLM backbone. We evaluated configurations with static analysis alone, RAG alone, and the combination of both.

As observed in Table \ref{tab:advanced_performance}, integrating external knowledge sources leads to notable improvements in {SmartAuditFlow}'s detection accuracy. Incorporating static analysis tools alone ({SmartAuditFlow} + static tool) increased the top-$max$ accuracy from 37.2\% to \textbf{39.9\%} (a +2.7 percentage point improvement). This enhancement can be attributed to the static analyzer's ability to flag determinate, pattern-based vulnerabilities and provide structural insights (e.g., identifying critical contract functions or inter-contract relationships) that refine the LLM's initial analysis.

Adding RAG alone ({SmartAuditFlow} + RAG) also improved accuracy, increasing it from 37.2\% to \textbf{40.1\%} (a +2.9 percentage point improvement). This underscores RAG's effectiveness in augmenting the LLM with external, up-to-date knowledge during the calibration phase, thereby enhancing its vulnerability verification capabilities and reducing potential hallucinations by grounding findings in retrieved evidence.
The most significant performance was achieved when both static analysis tools and RAG were integrated ({SmartAuditFlow} + Static Tool + RAG), yielding a top-$max$ accuracy of \textbf{41.2\%}. This represents a substantial +4.0 percentage point improvement over the baseline configuration. This result supports our hypothesis that leveraging multiple, complementary external knowledge sources synergistically enhances {SmartAuditFlow}'s overall detection performance.

The average number of outputs per contract remained remarkably stable and slightly decreased with enhanced knowledge integration (from 16.9 to 15.7). While ranking quality (MAP) showed some variation, the full integration of static tools and RAG notably improved this metric (from 0.253 to 0.281). This demonstrates that the accuracy gains from external knowledge do not lead to excessive output, and can concurrently improve ranking, ensuring auditor efficiency.

\begin{custommdframed}
  \textbf{Answer to RQ5:} The integration of external knowledge sources significantly enhances {SmartAuditFlow}'s performance. Incorporating both static analysis tools and RAG techniques boosts the top-$max$ detection accuracy by 4.0 percentage points (from 37.2\% to 41.2\%) over the baseline configuration. 
\end{custommdframed}

\begin{table*}[t]
  \centering
  \caption{Evaluation of SmartAuditFlow Performance with External Knowledge Integration}
  \small
  \label{tab:advanced_performance}
  \begin{threeparttable}
  \begin{tabular}{l | c | c | c }
    \toprule
    \textbf{Methods} & \textbf{Accuracy (top-max)} & \textbf{Avg. Outputs} & \textbf{MAP} \\
    \midrule
    SmartAuditFlow (Baseline) & 37.2\% & 16.9 & 0.253 \\
    SmartAuditFlow + static tool & 39.9\% (+2.7\%) & 16.1 & 0.242 \\
    SmartAuditFlow + RAG & 40.1\% (+2.9\%) & 16.7 & 0.257  \\
    SmartAuditFlow + static tool + RAG & 41.2\% (+4.0\%) & 15.7 & 0.281 \\
    \bottomrule
  \end{tabular}
  \end{threeparttable}
\end{table*}


\subsubsection{RQ6: Comparison with Other LLM-based Auditing Methods}
\label{subsubsec:rq6_other_llm_methods}
To contextualize {SmartAuditFlow}'s capabilities within the evolving landscape of LLM-assisted smart contract auditing, we compare its performance against several notable prior works.
Notably, David et al. \cite{david2023you}, PropertyGPT \cite{liu2024propertygpt}, and GPTScan \cite{sun2024gptscan} have demonstrated promising results using LLMs in different auditing scenarios. 

David et al. \cite{david2023you} employed direct prompting of GPT-4 and Claude-v1.3 for a binary classification task on 146 vulnerabilities across 38 types, identifying 58 TPs. PropertyGPT \cite{liu2024propertygpt} focused on verifying LLM outputs, correctly identifying 9 out of 13 CVEs (focused on 5 classic types) and 17 out of 24 vulnerabilities from their SmartInv dataset. GPTScan \cite{sun2024gptscan} targeted 10 common logic vulnerabilities, reporting 40 TPs. Their stated recall of 83.33\% suggests these 40 TPs were out of approximately 48 actual instances of those types within their specific evaluation context. In contrast, {SmartAuditFlow-Enhanced} was evaluated on the 784 high and medium severity vulnerability instances within our challenging Real-World Contracts Set, identifying 310 TPs. A key distinction of our work is the aim to address a broad spectrum of vulnerability types ("All Types"), rather than a predefined subset of common or logic vulnerabilities.

To facilitate a more direct comparison on a common ground, we evaluated {SmartAuditFlow-Enhanced} on the set of 13 representative CVEs previously analyzed by PropertyGPT \cite{liu2024propertygpt}. We compare our results with those reported for PropertyGPT, GPTScan, Slither, and Mythril (sourced from PropertyGPT's study), and further benchmark against standalone invocations of reasoning-enhanced LLMs: GPT-o3 and DeepSeek-R1. The detailed detection results are presented in Table \ref{tab:cve_comparison}.
{SmartAuditFlow-Enhanced} correctly detected all \textbf{13 out of 13 CVEs} in this benchmark, achieving a 100\% detection rate. This includes complex cases that other methods reportedly missed (e.g., CVE-2021-3004, CVE-2018-17111, as noted for PropertyGPT which detected 9/13).

Among other LLM-based approaches on this CVE set, standalone prompting of reasoning-enhanced models like GPT-o3 (using its specified version) and DeepSeek-R1 (version) also demonstrated strong capabilities, reportedly identifying 11/13 TPs and 9/13 TPs, respectively, in a single pass (pass@1).

\begin{custommdframed}
  \textbf{Answer to RQ6:}  When compared to other LLM-based auditing methods, {SmartAuditFlow-Enhanced} demonstrates a broader scope of vulnerability detection and achieved a high detection count (310 TPs) on a large set of real-world vulnerabilities. On a standardized 13-CVE benchmark, {SmartAuditFlow-Enhanced} achieved a 100\% detection rate, surpassing reported results for PropertyGPT (9/13), standalone reasoning LLMs like GPT-o3 (11/13), and traditional tools.
\end{custommdframed}

\begin{table*}[t]
  \centering
  \caption{Vulnerability Detection Performance on 13-CVE benchmark}
  \scriptsize
  \label{tab:cve_comparison}
  \begin{threeparttable}
  \begin{tabular}{l | c | c  c c c  c c c c}
    \toprule
    \textbf{CVE} & \textbf{Description}  & \textbf{Our work} & \textbf{David} & \textbf{PropertyGPT} & \textbf{GPTScan} & \textbf{Slither} & \textbf{Mythril} & \textbf{GPT-o3} & \textbf{Deepseek-r1}\\
    \midrule
    CVE-2021-34273  & access control  & \checkmark  &  $\times$ & \checkmark & \checkmark & $\times$ & $\times$ & \checkmark & \checkmark \\
    CVE-2021-33403 & overflow & \checkmark & $\times$ & \checkmark & $\times$ & $\times$ & \checkmark & \checkmark & \checkmark \\ 
    CVE-2018-18425 & logic error  & \checkmark & $\times$ & \checkmark & $\times$ & $\times$ & $\times$ & $\times$ & $\times$ \\ 
    CVE-2021-3004  & logic error  & \checkmark & $\times$ & $\times$ & $\times$ & $\times$ & $\times$ & $\times$ & $\times$ \\
    CVE-2018-14085   & delegatecall & \checkmark & \checkmark & $\times$ & $\times$ & \checkmark & $\times$ & \checkmark & \checkmark \\
    CVE-2018-14089   & logic error  & \checkmark & \checkmark & \checkmark & \checkmark & $\times$ &$\times$ &  \checkmark  & $\times$ \\
    CVE-2018-17111   & access control   & \checkmark & \checkmark & $\times$ & $\times$ & $\times$ & $\times$ &  \checkmark & \checkmark  \\
    CVE-2018-17987   & bad randomness   & \checkmark &\checkmark & $\times$ & \checkmark & $\times$ & $\times$ &  \checkmark & \checkmark \\
    CVE-2019-15079   & access control   & \checkmark & $\times$  & \checkmark & $\times$ & $\times$ & $\times$ &  \checkmark  & \checkmark \\
    CVE-2023-26488  & logic error  & \checkmark &  $\times$  & \checkmark & $\times$ & $\times$ & $\times$ & $\times$ &$\times$\\
    CVE-2021-34272   & access control   & \checkmark & $\times$  & \checkmark & \checkmark &$\times$ &$\times$ &  \checkmark & \checkmark \\
    CVE-2021-34270   & overflow  & \checkmark & \checkmark & \checkmark & \checkmark &$\times$ &\checkmark &  \checkmark & \checkmark \\
    CVE-2018-14087  & overflow  & \checkmark & $\times$ & \checkmark & $\times$ & $\times$&\checkmark &  \checkmark & \checkmark \\
    \bottomrule
  \end{tabular}
  \begin{tablenotes}
    \item Note: \checkmark\ indicates a correct detection (TP), whereas $\times$ indicates an incorrect detection (FP).
  \end{tablenotes}
  \end{threeparttable}
\end{table*}

\section{Related Work and Discussion}
\label{sec:related}
\subsection{Related Work}
Recent studies have demonstrated growing potential for applying LLMs to software vulnerability analysis and smart contract security. 
Liu et al. \cite{liu2024exploring} examined GPT's performance across six tasks—including bug report summarization, security bug report identification, and vulnerability severity evaluation. They compared GPT's results with 11 SOTA approaches. Their findings indicate that, when guided by prompt engineering, GPT can outperform many existing techniques. In parallel, Yin et al. \cite{yin2024multitask} established benchmark performance metrics for open-source LLMs across multiple vulnerability analysis dimensions. They emphasized how strategic prompt design and model fine-tuning substantially impact detection accuracy and localization precision.

Researchers have extended these investigations to domain-specific languages (DSLs) likeC/C++, Java, and Solidity.
Khare et al. \cite{khare2023understanding} conducted an comprehensive evaluation of 16 advanced LLMs in the context of C/C++ and Java, showing that LLM-based approaches can surpass traditional static analysis and deep learning tools. Li et al. \cite{li2024llm} integrated GPT-4 with CodeQL, a static analysis tool, to detect vulnerabilities in Java code, achieving an accuracy improvement of over 28\% compared to using CodeQL alone.

Building on these foundations, recent advancements employ auxiliary techniques to enhance LLM capabilities. LLM4Vuln \cite{sun2024llm4vuln} employs techniques such as knowledge retrieval, context supplementation, and advanced prompt schemes to boost vulnerability detection across multiple languages, including Java, C/C++, and Solidity. Their results show that GPT-4, when leveraging these enhancements, significantly outperforms competing models like Mixtral and CodeLlama. 
Similarly, GPTScan \cite{sun2024gptscan} demonstrates that a binary (Yes/No) response format can be effectively used by GPT models to confirm potential vulnerabilities in Solidity smart contracts. 

Furthermore, PropertyGPT \cite{liu2024propertygpt} integrates GPT-4 with symbolic execution at the source code level, enabling formal verification of smart contract properties. Meanwhile, Ma et al. \cite{ma2024combining}  propose a paradigm-shifting architecture that decouples detection and reasoning tasks through two-stage fine-tuning. Simultaneously, they improved both accuracy and model interpretability in smart contract analysis.

\subsection{Summary of Findings}

Based on the above evaluation results, we have derived the several key findings:

\begin{itemize}
  \item \textbf{Superior Vulnerability Detection through a Structured Workflow:} The core innovation of our methodology lies in decomposing the complex audit process into a Plan-Execute paradigm with distinct analytical stages. This structured workflow, mimicking a human auditor's systematic approach, enables the integrated LLM to perform a more thorough and focused analysis of specific code aspects or vulnerability types within each sub-task. This highlights the power of a well-designed workflow in enhancing detection capabilities.
  \item \textbf{Prompt Optimization and LLM Guidance:} The iterative prompt optimization strategy detailed in our Plan Phase is a foundational element contributing to {SmartAuditFlow}'s effectiveness. By systematically generating and selecting optimized prompts for critical stages like Initial Analysis (A1) and Audit Planning (A2), we ensure the LLM's reasoning is precisely guided towards relevant contract features and potential risk areas. By carefully crafting the prompt, our approach effectively handles ambiguous or complex code segments, leading to improved detection of subtle vulnerabilities. 
  \item \textbf{Effective Integration of External Knowledge Integration:} The integration of external knowledge sources demonstrably enhances {SmartAuditFlow}'s performance. Incorporating static analysis tool outputs provides crucial structural and pattern-based insights during the initial analysis (A1), while RAG enriches the calibration step (A3) with up-to-date, context-specific information. This underscores the value of a multi-source, hybrid approach.
  \item \textbf{Modularity and Flexibility with Play-and-Plugin Architecture:} {SmartAuditFlow}'s design exhibits significant modularity and flexibility. The framework's compatibility with various leading LLM backbones (e.g., Gemini, GPT, Claude series) allows users to select a model that best aligns with their specific performance priorities, such as maximizing detection accuracy, optimizing ranking quality (MRR), ensuring report conciseness (average outputs), or managing computational effort (average inference steps). 
  \item \textbf{Effectiveness on Complex and Edge-Case Scenarios:} {SmartAuditFlow} demonstrates robust performance not only on standard benchmarks but also on challenging real-world contracts and specific CVEs, which often represent edge cases or complex interaction-dependent vulnerabilities. Its ability to achieve high detection rates in these scenarios indicates its capacity to handle intricacies that often elude traditional tools and simpler LLM applications.
\end{itemize}

\subsection{Threats of Validity}

Our proposed system has the following potential limitations:
\begin{itemize} 
  \item \textit{Prompt Limitations for LLM:} Our iterative prompt optimization process, while automated, may yield prompts that are highly tuned to the specific LLM family. While RQ4 shows {SmartAuditFlow} functions with various LLMs using the same core prompts, its peak performance with a new LLM family might require re-running or adapting the prompt optimization. Further research into creating universally robust prompts or lightweight LLM-specific tuning of prompts is warranted.
  \item \textit{Computational Resources and Cost:} The multi-stage, iterative nature of {SmartAuditFlow}, involving multiple LLM calls per contract (about 30-40 steps in RQ4) and potentially intensive prompt optimization, demands considerable computational resources. Our experiments indicate analysis costs ranging from approximately \$0.1 USD for simple contracts to upwards of \$1 USD for highly complex ones using commercial APIs. While these costs are substantially lower than traditional human-led audits (often thousands of USD), they represent a consideration for widespread adoption, especially for batch processing of numerous contracts.
  \item \textit{Dependency on External LLM APIs}: The current implementation's reliance on commercial LLM APIs (e.g., for OpenAI, Google, Anthropic models) introduces external dependencies. Potential API changes, service disruptions, evolving pricing models, rate limits, and data privacy considerations could impact the system's long-term stability, operational reliability, and cost-effectiveness. Exploring hybrid models incorporating capable open-source, locally deployable LLMs could mitigate some of these external dependencies.
  \item \textit{Static Analysis Limitations}: {SmartAuditFlow} primarily performs a static analysis of smart contract code—both through its LLM-driven interpretation and its integration of static analysis tools. This approach is inherently limited in detecting vulnerabilities that only manifest during runtime or depend on complex state interactions specific to a live blockchain environment. 
\end{itemize}

\section{Conclusion}
\label{sec:conclusion}

This paper introduces SmartAuditFlow, a novel Plan-Execute framework for smart contract security analysis that leverages dynamic audit planning and structured execution. Unlike conventional LLM-based auditing methods that follow fixed, predefined steps, SmartAuditFlow dynamically adapts in real-time, generating and refining audit plans based on the unique characteristics of each contract. Its step-by-step execution process enhances vulnerability detection accuracy while minimizing false positives. The framework integrates iterative prompt optimization and external knowledge sources, such as static analysis tools and Retrieval-Augmented Generation (RAG), ensuring security audits are grounded in both code semantics and real-world security intelligence. Extensive evaluations across multiple benchmarks demonstrate SmartAuditFlow’s superior performance, achieving 97.2\% accuracy on common vulnerabilities, high precision in real-world projects, and the successful identification of all 13 tested CVEs.

Potential future enhancements include integrating knowledge graphs into the RAG process, adopting formal verification techniques to improve LLM output accuracy, and optimizing prompt strategies for increased efficiency. Additionally, improving audit decision interpretability will be essential for real-world adoption. We envision SmartAuditFlow as a significant advancement in automated smart contract auditing, offering a scalable, adaptive, and high-precision solution that enhances blockchain security and reliability.

\bibliographystyle{unsrtnat}
\bibliography{references} 

\appendix

\section{Prompt Optimization for Initial Analysis}
\label{sec:prompt_initial}

\paragraph{Objectives:} To identify an optimal instruction $\rho^*$ that, when provided to the LLM (gpt-4o-2024-11-20), maximizes the quality of smart contract analysis focusing on function definitions, state variables, modifiers, and events, according to the objective function in Eq. \ref{eq:optimization}. That instruction will be used in Step A1.

\subsection{Problem Setup}

\paragraph{1. Dataset Preparation:} We curated a high-quality dataset of 1,000 smart contract-analysis pairs, reflecting 130+ typical vulnerability types and 200 scenarios. Aim for publicly available, audited contracts where possible (e.g., from OpenZeppelin, well-known DeFi projects, Etherscan verified sources).
The dataset was carefully constructed to ensure comprehensive coverage and alignment with industry audit standards, as shown in Table \ref{tab:dataset_statistics}. 

For each contract $c$ and "expexted result" $A$ in $D_{task}$. This is the most critical and labor-intensive part. Human smart contract auditors and experienced developers will manually create a detailed analysis $A$. This analysis must explicitly:

\begin{itemize}
  \item List all function definitions, including their name, visibility, parameters (with types), return values (with types), and a concise description of their purpose and core logic.
  \item List all state variables, including their name, type, visibility, and a brief description of their purpose.
  \item List all modifiers, including their name, parameters (if any), and a description of the conditions they enforce or checks they perform.
  \item List all events, including their name, parameters, and a description of when and why they are emitted.
  \item A brief summary of how these components interact or potential questions for specific use of contracts.
\end{itemize}

\begin{table*}[h]
  \centering
  \caption{Prompt Optimization Dataset Statistics}
  \small
  \label{tab:dataset_statistics}
  \begin{threeparttable}
  \begin{tabular}{l | c | c  }
    \toprule
    \textbf{Metric} & \textbf{Training Set} & \textbf{Validation Set} \\
    \midrule
    Low Complexity & 400 & 100  \\ 
    Medium Complexity & 240 & 60  \\ 
    High Complexity & 160 & 40  \\ \midrule
    Total Contracts & 800 & 200  \\ 
    Vulnerability Types & 130 & 130  \\ 
    Avg. Vulnerabilities & 1.5 per & 1.5 per  \\ 
    \bottomrule
  \end{tabular}
  \end{threeparttable}
\end{table*}

\paragraph{2. Complexity Metric (Complexity($\rho$)):} 
\[
Complexity(\rho)= \omega_1 \cdot TokenCount(\rho) + \omega_2 \cdot SentenceCount(\rho)
\]
where $\omega_1 = 0.7$ and $\omega_2 = 0.3$. $TokenCount(\rho)$ is the number of tokens in the instruction $\rho$, and $SentenceCount(\rho)$ is the number of sentences in the instruction $\rho$.

\paragraph{3. Instruction Similarity Metric ($sim(\rho, \rho')$):} 
Use sentence embeddings (from a Sentence-BERT model like all-MiniLM-L6-v2) to represent $\rho$ and $\rho'$, then calculate their cosine similarity.

\paragraph{4. Multi-Criteria Scoring Function \(f(\rho, c, A)\):}   
\[
  f(\rho, c, A) = w_{\text{exec}} \cdot f_{\text{exec}}(\mathcal{M}(\rho, c), A) + w_{\text{log}} \cdot f_{\text{log}}(\mathcal{M}(\rho, c), A)
\]  
where $w_{\text{exec}}$ is 0.7 and $w_{\text{log}}$ is 0.3.

\paragraph{(1) Output Alignment Score $f_{\text{exec}}(\mathcal{M}(\rho, c))$:} 
Let $O=\mathcal{M}(\rho, c)$ be the LLM's generated analysis. We will parse both $O$ and the expert analysis A to extract structured information about:
\begin{itemize}
  \item $S_F$: Set of identified functions (name, params)
  \item $S_V$: Set of identified state variables (name, type)
  \item $S_M$: Set of identified modifiers (name)
  \item $S_E$: Set of identified events (name, params)
  \item $S_Q$: Set of additional questions or situations.
\end{itemize}

\[
  f_{\text{exec}}(\mathcal{M}(\rho, c)) = w_{\text{cov}} \cdot f_{\text{coverage}}(\mathcal{M}(\rho, c), A) + w_{\text{det}} \cdot f_{\text{detail}}(\mathcal{M}(\rho, c), A)
\] 

\paragraph{(2) Component Coverage Coverage ($f_{coverage}$):} Ensure the LLM output $\mathcal{M}(\rho, c)$ identifies all relevant structural components (functions, state variables, modifiers, events) present in the smart contract $c$ and doesn't list components that don't exist.

For each component type (e.g., functions, variables), compare the extracted set from $\mathcal{M}(\rho, c)$ ($S_F(O)$) with the set from $A$ ($S_F(A)$). Calculate Recall by comparing the sets extracted from $O$ with those from $A$.
\begin{itemize}
  \item $P_F = |S_F(O)\cap S_F(A)|/|S_F(O)|$, how many were correct in the number of the LLM output?
  \item $R_F = |S_F(O)\cap S_F(A)|/|S_F(A)|$, how many were correct in all correct items?
  \item $F1_{F} = 2 \cdot P_F \cdot R_F / (P_F + R_F)$
\end{itemize}
Finally, $f_{coverage}$ can be a weighted average of the F1-scores for each component type (e.g., functions might be weighted higher than events if deemed more critical for the analysis). $f_{coverage} = w_{\text{F}} \cdot F1_{F} + w_{\text{V}} \cdot F1_{V} + w_{\text{M}} \cdot F1_{M} + w_{\text{E}} \cdot F1_{E}$.

\paragraph{(3) Detail Accuracy Score (($f_{detail}$)):} For each component correctly identified (i.e., true positives from the $f_{coverage}$ step), assess the accuracy and completeness of its description in $\mathcal{M}(\rho, c)$ compared to the detailed description in $A$. Inspired by RAGAS \cite{es2024ragas}, use another powerful LLM (the "judge model") to evaluate the quality of the primary LLM's description against the expert analysis $A$. The judge model would be prompted with the primary LLM's description of a component (e.g., function foo), the expert description of foo from $A$, and a rubric. 

The rubric ask the judge to score aspects like:
\begin{itemize}
  \item Factual Correctness: (e.g., "Does the LLM accurately state the visibility of function foo as per the expert analysis?") - Scale of 1-5
  \item Completeness of Detail: (e.g., "Does the LLM mention all parameters of function foo listed in the expert analysis?") - Scale of 1-5 or Yes/No.
  \item Relevance of Information: (e.g., "Is the LLM's description of foo's purpose relevant and free of unnecessary detail when compared to the expert summary?")
\end{itemize}

Output Likelihood Score $f_{\text{log}}(\mathcal{M}(\rho, c), A)$: Calculated as per Eq. \ref{eq:log_likelihood} in Section \ref{subsubsec:multi_criteria_scoring} (average negative log-likelihood per token of the LLM's output).

\subsection{Procedure}

\paragraph{1. Initial Instructions Set $\mathcal{U}_0$} 
Set initial population size as $k=20$. Stochastic sampling for Stage A1:

Meta-Prompt: \textit{You are an expert prompt engineer. Generate 20 diverse candidate instructions for an LLM. The LLM's task is to produce a precise and comprehensive analysis of a given smart contract. The analysis must specifically and accurately detail: 1. All function definitions (including name, visibility, parameters, return values, and purpose). 2. All state variables (including name, type, visibility, and purpose). 3. All modifiers (including name, parameters, and conditions enforced). 4. All events (including name, parameters, and emission context). The instructions should be clear, unambiguous, and encourage thoroughness.}

\paragraph{Mutation of Seed Prompts}
Manually crafted seed prompts representing different approaches, for example:
\begin{itemize}
\item $\rho_{seed1}$: "Analyze the provided smart contract. Focus on its function definitions, state variables, modifiers, and events. Be precise."
\item $\rho_{seed2}$: "List all functions, state variables, modifiers, and events in the smart contract. For each function, detail its parameters, return values, and purpose. For state variables, list type and purpose. For modifiers, explain their checks. For events, list parameters and when they are emitted."
\end{itemize}

Apply paraphrasing, keyword substitution (e.g., "list" vs "detail" vs "enumerate"; "purpose" vs "functionality"), and structural edits (reordering points, changing from imperative to declarative). Initial temperature $\tau = 0.7$ for mutation, and replay buffer (\(\mathcal{B}_{\text{replay}}\)) as empty.

\paragraph{2. Evolution Loop}
Max generations: $T = 10$ (max generations).

Mini-Batch sampling: $n_t = \lceil 0.1 \cdot |\mathcal{D}_{train}| \cdot (1 + t/T) \rceil$. Since $|\mathcal{D}_{train}|=800$, $n_t$ starts at 80 and goes up to 160.

Stochastic fitness evaluation: Use Eq. \ref{eq:fintness} with $\epsilon = 0.1$ (weight for replay-based regularization).

Elite selection with momentum: $k_e = 10$ (number of elites). $\alpha = 0.3$ (momentum coefficient).

Offspring generation with guided mutation: Generate $k - k_e = 10$ offspring.
Set $\tau_{\text{initial}} = 0.7$, $\beta = 0.1$ for exponential decay of mutation temperature ($\tau_t = \tau_{\text{initial}} \cdot e^{-\beta t}$).Prioritize mutations  that show improvement on the current mini-batch $\mathcal{B}_t$ for a subset of mutations.

Population Update: $\mathcal{U}_t = \mathcal{E}_t \cup \mathcal{O}_t$. Update $\mathcal{B}_{\text{replay}}$ with the top $k_e$ instructions from $\mathcal{U}_t$ and their $\bar{f}_t(\rho)$ scores if they are better than what's stored.

Convergence Check:
\begin{itemize}
  \item Terminate if the moving average fitness $\bar{f}_t(\rho)$ of the top elite does not improve by more than $\delta = 0.005$ for 5 consecutive generations.
  \item Terminate if population diversity $D(\mathcal{U}_t)$ (average pairwise similarity) falls below, e.g., 0.3 (if instructions become too similar).
\end{itemize}

\paragraph{3. Final Evaluation}
Upon termination, take all instructions from the final population $\mathcal{U}_T$.

Evaluate each $\rho \in \mathcal{U}_T$ using the full objective function (Eq. \ref{eq:optimization}) on the held-out validation set $\mathcal{D}_{\text{val}}$. Use regularization hyperparameter $\lambda = 0.01$ for complexity; It needs tuning, could be another parameter sweep.

The $\rho^*$ that maximizes this score on $\mathcal{D}_{\text{val}}$ is selected as the optimal instruction.

\paragraph{4. Hyperparameter Summary (Initial Values)}:
Population size ($k$): 20; Elite size ($k_e$): 10; Max generations ($T$): 10;
Initial mutation temperature ($\tau_{\text{initial}}$): 0.7; Mutation temperature decay rate ($\beta$): 0.3; Replay regularization weight ($\epsilon$): 0.1; Momentum coefficient ($\alpha$): 0.3; Complexity regularization ($\lambda$): 0.01; Mini-batch base fraction: 0.1; Fitness improvement threshold ($\delta$): 0.005; Scoring weights: $w_{\text{exec}}=0.7, w_{\text{log}}=0.3$; (within $f_{exec}$) $w_{cov}=0.6, w_{det}=0.4$.

\paragraph{5. Evaluation of the Experiment:}
\begin{itemize}
  \item The primary output is the optimal instruction $\rho^*$.
  \item Secondary evaluations could include: Tracking the average and best fitness over generations to observe convergence; Analyzing the characteristics of $\rho^*$ (length, specific keywords, structure); Qualitative human evaluation of analyses generated by $\rho^*$ on a few unseen contracts compared to analyses from seed prompts or baseline prompts.
\end{itemize}

\paragraph{Expected Outcome}:
The final optimal instruction $\rho^*$ is list in our Guithub repositories. We present an examplre of initial instruction set scoring in Table \ref{tab:p0_result}.

\begin{table*}[h]
  \centering
  \caption{Prompt Optimization Dataset Statistics}
  \small
  \label{tab:p0_result}
  \begin{threeparttable}
    \begin{tabular}{l|l|c|c|c}
    \toprule
    \textbf{Temp.} &\textbf{Instruction} & \textbf{Execution Score} & \textbf{Log Score (Mean)} & \textbf{Combine Score}  \\
    \midrule
    0.1 & $\rho_0$ & 95.828 & -88.628  & \textbf{40.491} \\ 
    0.1 & $\rho_1$ & 96.234 & -103.886 & \textbf{36.198} \\  
    0.1 & $\rho_2$ & 96.984 & -90.116 & \textbf{40.854} \\ 
    0.1 & $\rho_3$ & 94.718 & -105.129 & 34.764 \\ 
    0.1 & $\rho_4$ & 96.056& -109.270 & 34.459\\ 
    0.4 & $\rho_5$ & 79.779 & -74.885 & 33.379 \\ 
    0.4 & $\rho_6$ & 80.375 & -73.798 & 34.122 \\ 
    0.4 & $\rho_7$ & 70.532 & -108.916 & 16.697\\ 
    0.4 & $\rho_8$ & 75.615 & -43.416 & \textbf{39.905}\\ 
    0.4 & $\rho_9$ & 69.578 & -46.807 & 34.662\\ 
    0.7 & $\rho_{10}$ & 68.710 & -52.154 & 32.451 \\ 
    0.7 & $\rho_{11}$ & 71.546 & -52.366 & 34.372\\ 
    0.7 & $\rho_{12}$ & 75.234 & -60.393 & 34.546\\ 
    0.7 & $\rho_{13}$ & 70.875 & -47.927 & 35.234\\ 
    0.7 & $\rho_{14}$ & 77.839 & -71.377 & 33.074 \\ 
    1.0 & $\rho_{15}$ & 70.662 & -60.360 & 31.355\\ 
    1.0 & $\rho_{16}$ & 69.232 & -69.027 & 27.754\\ 
    1.0 & $\rho_{17}$ & 72.078 & -52.578 & 34.681 \\ 
    1.0 & $\rho_{18}$ & 69.375 & -13877.186 & -4114.593\\ 
    1.0 & $\rho_{19}$ & 71.399 & -11339.222 & -3351.787 \\ 
    \bottomrule
  \end{tabular}
  \end{threeparttable}
\end{table*}

\section{Retrievable Domain Knowledge Base}
\label{sec:rag_baseknowledge}
\textbf{Goal:} To construct a comprehensive, accurate, up-to-date, and efficiently retrievable knowledge base focused on smart contract security, vulnerabilities, best practices, and exploits, serving as the foundation for our Retrieval-Augmented Generation (RAG) system.

\subsection{Foundation and Planning}
The scope and objectives of the knowledge base were clearly defined, and a plan was developed to ensure its success. The primary scope and objectives include:

\begin{itemize}
  \item Blockchain Platforms: Initially focus on Ethereum, then potentially expand for future. 
  \item Smart Contract Languages: Primarily Solidity, as it is the most prevalent language for smart contracts.
  \item Vulnerability Categories: Prioritization of common and critical vulnerabilities (e.g., Reentrancy, Integer Over/Underflow, Access Control, Unchecked External Calls, Gas Limit Issues), while aiming for comprehensive coverage of known smart contract weaknesses.
  \item {Knowledge Types:} A diverse range of information sources are targeted:
    \begin{itemize}
        \item Formal vulnerability definitions and classifications (e.g., SWC Registry, relevant CWEs).
        \item Official security guidelines and best practices from authoritative bodies (e.g., Ethereum Foundation, Consensys, OpenZeppelin).
        \item Seminal and contemporary research papers on smart contract security and analysis.
        \item Detailed technical blog posts, articles, and post-mortems dissecting real-world exploits and novel vulnerabilities from reputable security researchers and firms.
        \item Repositories of secure coding patterns and anti-patterns.
        \item Publicly available smart contract auditing reports and findings.
        \item Relevant sections of smart contract language documentation focusing on security considerations.
    \end{itemize}
\end{itemize}

\paragraph{Curation Criteria}
To ensure the quality, relevance, and reliability of the knowledge base, the following curation criteria were established:
\begin{itemize}
  \item Source Authority: Prioritization of official documentation, peer-reviewed academic papers, reports from reputable security firms (e.g., Trail of Bits, Sigma Prime, Consensys Diligence), and content from well-known, respected security researchers.
  \item Accuracy and Verifiability: Information must be accurate and, where possible, cross-verified with multiple sources. Claims should be supported by evidence or clear reasoning.
  \item Recency and Relevance: For exploit descriptions, emerging vulnerabilities, and evolving best practices, newer information is generally preferred. Foundational concepts may come from seminal works. All content must be directly relevant to smart contract security. 
  \item Clarity and Detail: Preference for sources that provide in-depth explanations, illustrative code examples (vulnerable and remediated), and actionable mitigation strategies.
\end{itemize}

\paragraph{Data Schema:} For each entry/document chunk into the knowledge base, the data schema is designed to capture the following key fields:
\begin{itemize}
  \item content: The actual text chunk.
  \item source\_url: The original URL or identifier.
  \item source\_type: (e.g., "SWC", "Blog", "Research Paper", "Guideline", "CVE").
  \item title: Original title of the document.
  \item publication\_date: (YYYY-MM-DD).
  \item last\_accessed\_date: When it was last retrieved/checked.
  \item vulnerability\_tags: (e.g., "reentrancy", "integer\_overflow", "access\_control").
  \item platform\_tags: (e.g., "ethereum", "solidity").
  \item severity\_keywords: (e.g., "critical", "high", "medium", "low",informational", "ground").
  \item chunk\_id: Identifier for the specific text chunk.
  \item document\_id: Identifier for the parent document.
  \item summary: A brief LLM-generated summary of the chunk.
\end{itemize}

\subsection{Content Acquisition and Ingestion}
The process of acquiring and ingesting content involves systematic collection from diverse sources, followed by meticulous extraction and cleaning.

\paragraph{Source Collection:} A curated list of target websites, databases, academic journals, and code repositories was compiled based on the established curation criteria. Examples of these sources are provided in Table~\ref{tab:domain_knowledge_examples}. Content acquisition is achieved through:

\begin{itemize}
  \item {Automated Methods:} Utilizing web scraping tools (Python libraries: BeautifulSoup) and APIs to systematically gather information from predefined sources.
  \item {Manual Curation:} Supplementing automated collection by manually gathering specific high-value documents, such as PDFs of research papers or audit reports, especially from sources where scraping is not feasible or permitted.
\end{itemize}

\paragraph{Data extraction and cleaning:}
Raw content extracted from various sources undergoes a rigorous cleaning process:
\begin{itemize}
  \item {Text Extraction:} Conversion of diverse formats (HTML, PDF, Markdown) into plain text.
  \item {Normalization:} Standardizing whitespace, character encodings, and resolving minor formatting inconsistencies to prepare the text for further processing.
\end{itemize}
The cleaned and structured data is then managed through a data pipeline for ingestion into the knowledge base.

\begin{table*}[ht]
  \centering
  \scriptsize
  \caption{Illustrative Examples of Domain Knowledge Sources for Smart Contracts}
  \label{tab:domain_knowledge_examples}
  \begin{threeparttable}
  \begin{tabular}{p{0.12\textwidth}|p{0.13\textwidth}|p{0.25\textwidth}|p{0.38\textwidth}}
      \toprule
      \textbf{Category} & \textbf{Source Name} & \textbf{Description/Relevance} & \textbf{Example URL/Identifier} \\
      \midrule
      Official Docs & Solidity Language & Security considerations, language specifics & \url{docs.soliditylang.org/en/latest/security-considerations.html} \\
      Vulnerability DB & SWC Registry & Smart Contract Weakness Classification & \url{securing.github.io/SCSVS/} \\
      Vulnerability DB & DASP-Top-10 (OWASP) & Top 10 vulnerabilities (older but foundational) & \url{www.dasp.co/} \\
      Security Guidelines & Consensys & Known attacks, best practices & \url{consensys.io/diligence/blog} \\
      Security Guidelines & OpenZeppelin & Secure development guides, contract standards & \url{openzeppelin.com/contracts} \\
      Research & Academic Papers & Peer-reviewed security analyses, new techniques & \tiny{arXiv, IEEE, ACM Digital Library} \\
      Technical Blogs & Trail of Bits Blog & In-depth vulnerability analysis, security tools & \url{blog.trailofbits.com} \\
      Technical Blogs & Sigma Prime Blog & Ethereum security, client vulnerabilities & \url{blog.sigmaprime.io} \\
      Community Standard & SCSVS & Smart Contract Security Verification Standard & \url{securing.github.io/SCSVS/} \\
      ecurity Guidelines & 2024 Web3 Security Report & Smart Contract Auditing Tools Review  & \url{https://hacken.io/discover/audit-tools-review/} \\
      Audit Reports & Various Firms & Public reports from reputable auditors & \tiny{(CertiK, PeckShield, Code4rena, SlowMist,)} \\
      \bottomrule
  \end{tabular}
  \end{threeparttable}
\end{table*}

\subsection{Processing, Chunking, and Embedding}
Once cleaned, the textual data is processed to optimize it for retrieval by the RAG system.

\paragraph{Document Parsing:} The cleaned text is parsed to identify its internal structure, such as headings, paragraphs, lists, and code blocks. Special attention is given to code snippets to preserve their formatting and potentially identify the programming language.

\paragraph{Content Chunking:} Effective chunking is critical for RAG performance. Our primary approach is {semantic chunking}, which groups text based on topical coherence. This is implemented using techniques available in libraries (such as `Langchain'), aiming to create chunks that are topically focused yet contextually rich.
Chunks are designed to be small enough for efficient processing by the embedding model and for focused retrieval, yet large enough to encapsulate meaningful semantic units.
As a fallback or for specific document types, fixed-size chunking with a defined token overlap (e.g., 256-512 tokens per chunk with 10-20\% overlap) may be utilized to ensure complete coverage while maintaining local context.

\paragraph{Embedding Generation}
For each processed text chunk, a dense vector embedding is generated using a pre-trained sentence transformer model (embedding models suitable for semantic similarity tasks). These embeddings capture the semantic meaning of the text chunks.

\paragraph{Storage and Indexing}
The text chunks, along with their corresponding vector embeddings and associated metadata (as defined in the data schema), are ingested and stored in a specialized Vector Database (Milvus).
The Vector Database is configured with an appropriate indexing strategy (HNSW) to enable efficient and scalable approximate nearest neighbor (ANN) search. This allows for rapid retrieval of the most semantically similar document chunks in response to a query. Hybrid search capabilities, combining semantic similarity with keyword-based filtering on metadata, are also leveraged where available.

\begin{table*}[ht]
  \centering
  \scriptsize
  \caption{The Key Components of the SmartAuditFlow Framework}
  \label{tab:algorithm_overview}
    \begin{threeparttable}
  \begin{tabular}{|c|p{2cm}|p{4cm}|p{4cm}|p{4cm}|}
      \hline
      \textbf{Step} & \textbf{Step Name} & \textbf{Objective} & \textbf{Inputs} & \textbf{Outputs} \\
      \hline
      1 & \textbf{Initial Analysis} & 
      Conduct a preliminary examination of the smart contract to understand its structure and identify critical components. & 
      - Smart contract code ($c$) \newline 
      - Static analysis tool outputs (optional) & 
      - Structural representation of the contract \newline 
      - Identification of critical components \newline 
      - Summary of potential areas of concern \\
      \hline
      2 & \textbf{Audit Planning} & 
      Develop a systematic audit strategy by decomposing the audit into manageable sub-tasks, prioritizing high-risk areas. & 
      - Smart contract code ($c$) \newline 
      - Initial analysis results ($s_1$) \newline 
      - Risk assessment data & 
      - Set of audit sub-tasks ($\mathbf{t} = \{t^1, t^2, \dots, t^n\}$) \newline 
      - Prioritized audit plan $s_2$ \\
      \hline
      3 & \textbf{Multi-task Execution} & 
      Execute and validate each specialized security analysis sub-task to identify vulnerabilities and ensure accuracy. & 
      - Smart contract code ($c$) \newline 
      - Audit sub-tasks ($\mathbf{t}$) \newline 
      - Execution functions ($f_e^i$) \newline 
      - Validation functions ($f_v^i$) & 
      - Execution results for each sub-task ($\{e^1, e^2, \dots, e^n\}$) \newline 
      - Validation results for each sub-task ($\{v^1, v^2, \dots, v^n\}$) \newline 
      - Verified vulnerability findings \\
      \hline
      4 & \textbf{Findings Synthesis} & 
      Integrate and correlate results from all sub-tasks to assess the combined impact on contract security. & 
      - Verified vulnerability findings ($\{v_3^i\}$) \newline 
      - Structural and component insights ($s_1$) \newline 
      - Audit plan ($s_2$) & 
      - Correlated vulnerability map \newline 
      - Combined impact assessment \newline 
      - Systemic issue identification \\
      \hline
      5 & \textbf{Comprehensive Report} & 
      Generate a detailed audit report compiling all findings, providing remediation recommendations and overall security assessment. & 
      - Initial analysis ($s_1$) \newline 
      - Audit plan ($s_2$) \newline 
      - Findings synthesis ($s_4$) & 
      - Detailed audit report \newline 
      - Executive summary \newline 
      - Risk ratings and mitigation strategies \newline 
      - Compliance checks \\
      \hline
  \end{tabular}
\end{threeparttable}
\end{table*}





\end{document}